\documentclass[twocolumn]{aastex631}
\usepackage{amsmath,mathtools,mathrsfs}
\usepackage[caption=false]{subfig}
\usepackage[T1]{fontenc}
\usepackage{hyperref}
\begin{document}
\title{Principal-Component Interferometric Modeling (\texttt{PRIMO}), an Algorithm for EHT Data I: Reconstructing Images from Simulated EHT Observations}

\author{Lia Medeiros}
\altaffiliation{NSF Astronomy and Astrophysics Postdoctoral Fellow}
\affiliation{School of Natural Sciences, Institute for Advanced Study, 1 Einstein Drive, Princeton, NJ 08540}
\author{Dimitrios Psaltis}
\affiliation{Steward Observatory and Department of Astronomy, University of Arizona, 933 N. Cherry Ave., Tucson, AZ 85721}
\author{Tod R. Lauer}
\affiliation{NSF's National Optical Infrared Astronomy Research Laboratory, Tucson, AZ 85726}
\author{Feryal {{\"O}zel}}
\affiliation{Steward Observatory and Department of Astronomy, University of Arizona, 933 N. Cherry Ave., Tucson, AZ 85721}

\begin{abstract}
The sparse interferometric coverage of the Event Horizon Telescope (EHT) poses a significant challenge for both reconstruction and model fitting of black-hole images. \texttt{PRIMO} is a new principal components analysis-based algorithm for image reconstruction that uses the results of high-fidelity general relativistic, magnetohydrodynamic simulations of low-luminosity accretion flows as a training set. This allows the reconstruction of images that are both consistent with the interferometric data and that live in the space of images that is spanned by the simulations. \texttt{PRIMO} follows Monte Carlo Markov Chains to fit a linear combination of principal components derived from an ensemble of simulated images to interferometric data. We show that \texttt{PRIMO} can efficiently and accurately reconstruct synthetic EHT data sets for several simulated images, even when the simulation parameters are significantly different from those of the image ensemble that was used to generate the principal components. The resulting reconstructions achieve resolution that is consistent with the performance of the array and do not introduce significant biases in image features such as the diameter of the ring of emission.

\end{abstract}

 
\section{Introduction}\label{sec:intro}

The Event Horizon Telescope (EHT) collaboration recently imaged the supermassive black hole in the nearby giant elliptical galaxy M87 for the first time using sub-mm VLBI observations \citep{2019ApJ...875L...1E, 2019ApJ...875L...2E, 2019ApJ...875L...3E, 2019ApJ...875L...4E, 2019ApJ...875L...5E, 2019ApJ...875L...6E}. The first polarized images of the black hole in M87 were published a short time later and indicated a strong and ordered magnetic field in the vicinity of the black hole \citep{2021ApJ...910L..12E,2021ApJ...910L..13E}. 

Reconstructing images of the M87 supermassive black hole was challenging. The 2017 observations included only five telescope locations, resulting in markedly sparse interferometric ($uv$-plane) coverage. This challenge was extensively addressed in the EHT papers and particularly in \citet{2019ApJ...875L...4E}, which is mainly concerned with a detailed discussion of the image reconstruction techniques used. In brief, a variety of algorithms was employed and all were extensively tested with simulations and inter-compared on the images recovered from the actual observations.  Of necessity, each algorithm incorporated a variety of assumptions to address the incomplete $uv$-plane coverage, which in turn imply associated uncertainties in the images recovered. The aim of this diverse approach was to be conservative with the reconstructions and ensure that the major quantities of astrophysical interest that were recovered from the images were robust.

We begin with a discussion of the general image reconstruction techniques used so far, followed by the motivation for the \texttt{PRIMO} methodology that we introduce here.

\smallskip

\noindent \textit{General purpose imaging algorithms:} These include the traditional CLEAN algorithm \citep{1974A&AS...15..417H}, as well as new maximum likelihood methods (see e.g., \citealt{2019ApJ...875L...4E, 2016ApJ...829...11C,2017AJ....153..159A}). The challenge for general-purpose image reconstruction algorithms is to generate an image among an infinite set of formally allowable solutions that are compatible with the data. In order to reduce the range of possible solutions, regularizers and secondary constraints (such as image global entropy, smoothness, local curvature, etc.) are levied to recover an image that matches expectations of realistic structure.  These methods are agnostic to theoretical predictions on image morphology and can therefore be used to determine basic image features such as the presence of a ring or brightness depression. However, introducing constraints on the plausibility of the image components is unavoidable and can lead to artifacts as shown, e.g., in Figure 10 of \citet{2019ApJ...875L...4E}. Moreover, even though the regularizing conditions are reasonable for some astronomical images, they may not be well motivated for black hole images since simulations predict steep gradients in parts of the image~\citep{2015ApJ...814..115P}.

\smallskip 

\noindent \textit{Geometric fits:} These are posterior sampling algorithms that fit semi-analytic or geometric crescent- and ring-like models directly to interferometric data \citep{2013MNRAS.434..765K,2019ApJ...875L...6E}. The models invoke a much smaller number of free parameters and, therefore, do not require additional regularizers the way that the general purpose imaging algorithms do, as described above. However, in some cases these simple models may not be able to reproduce the complex image morphology predicted for black hole images. Indeed, simulations predict that the turbulent flows generate complex and stochastic structures as a consequence of the presence of bright, magnetically dominated flux tubes that are lensed by the black hole (see e.g., \citealt{2015ApJ...812..103C,2019ApJ...875L...5E}). Since the expected level of complexity is not included in the geometric model fits, the posteriors of the model parameters are affected by the most influential data points and may be biased \citep{2022ApJ...928...55P}.

\smallskip

\noindent \textit{Comparisons to numerical simulations:} These methods compare simulated images from general relativistic magnetohydrodynamic (GRMHD) simulations to interferometric data allowing for a rotation and scaling of the image relative to the data (see e.g., \citealt{2019ApJ...875L...5E}). This comparison leads to constraints on physically meaningful parameters about the accretion flow. However, a single EHT observation corresponds to a particular realization of the turbulent structure of the accretion that may be consistent with simulations only in a statistical sense. As a consequence, these methods benefit from prior characterization of the statistics of the various image structures and of the corresponding interferometric observables~\citep{2016ApJ...832..156K,2019ApJ...875L...6E}.

We present a novel principal-component interferometric modeling (\texttt{PRIMO}) algorithm that combines the desirable characteristics of the methods listed above while attempting to reduce their limitations. \texttt{PRIMO} uses a large library of GRMHD simulations as a ``training set'' for image reconstruction and model fitting. Instead of employing images that are smooth (as in the case of the maximum likelihood imaging methods) or consist of a limited number of broadened point sources (as in the case of CLEAN), it utilizes images that are broadly consistent with the space of possibilities spanned by the simulations. Because it involves a relatively small number of parameters, i.e., the coefficients of the principal components, it does not require imposing regularizers as is done in maximum likelihood methods. Furthermore, it is not limited to simple geometric shapes, such as crescents and rings, and can accurately reconstruct the stochastic features expected in black hole images. At the same time, it does not compare specific realizations of the turbulent images with the data but rather uses a principle-component decomposition to derive a basis for the space of possible images that are consistent with theoretical expectations. Finally, the PCA algorithm provides not only the best-fit image but rather a complete posterior over all image structures that are consistent with the data.

In addition to PCA, several other decompositions have been developed and applied to a multitude of problems. Using bases (called dictionaries), derived from PCA or other decompositions, to sparsely represent a training set falls under that umbrella of dictionary learning (see e.g. \citealt{Shao2014} for a review of dictionary learning applied to image de-noising). Within astronomy dictionary learning is frequently used to de-noise images and spectra, or for image classification. Convolutional neural networks (CNN) are also becoming ubiquitous in astronomy and have recently been applied to the output of the \texttt{Clean} algorithm to de-noise the results of image reconstructions \citep{2022MNRAS.509..990G}. Our goal is not de-noising in the image domain, \texttt{PRIMO} reconstructs images directly from the fourier-domain visibilities. PCA is well-suited for our application since it enables remarkably powerful dimensionality reduction, allowing us to fit only 20 PCA components to the visibilities. Nonnegative matrix factorization (NMF), for example, is also commonly used in astronomical applications (see e.g. \citealt{2016arXiv161206037Z}). However, requiring that the basis functions be positive definite can result in biases if the basis is truncated, especially near steep gradients like those expected near the boundary of the black hole shadow.

The PCA approach is very general but employs its own restrictions on the subset of allowable images by only requiring that the solution is likely to fall within the span of image morphologies produced by the training set of simulations. However, as it is well known~\citep[see e.g., ][]{10.1162/jocn.1991.3.1.71} and we will also demonstrate later, the PCA-based algorithm can reconstruct images even if the particular image structures are different in their details from the individual simulation snapshots that were used for the training set. Therefore, the method can be applied to reconstruct a black hole image even if the GRMHD outputs do not precisely represent all of its characteristics.

In \citet{2018ApJ...864....7M}, we showed that PCA could be used to efficiently represent the ``space'' of image morphologies seen in GRMHD simulations of an accreting black hole. The full range of structures seen in a simulation can then be encoded as a linear combination of a compact set of orthogonal ``eigenimages,''  with each eigenimage describing a portion of the structure seen in the simulation. Critically, PCA minimizes the number of components needed to describe the full variance of the simulation and the components can be ordered by the decreasing fraction of the variance that they describe.

A particular benefit of the PCA approach is that the orthogonal compact basis derived in image space transforms identically to the same basis that  would be derived directly by representing the simulations in visibility (Fourier) space (see \citealt{2018ApJ...864....7M} for a mathematical proof). In short, the basis can be built in the image domain, where we have the best {\it a priori} knowledge of the likely image morphology, but is fitted in the complementary visibility space in which the observations are presented.

Another benefit of \texttt{PRIMO} is that it not only provides excellent recovery of structure up to the formal resolution limit of the observations, but can provide ``super-resolution'' at yet finer scales. Rich knowledge of the intrinsic source structure allows for quantitative measures of features that could not be recovered without strong priors. The principal-component basis encodes the intrinsic correlations of the source structure over a range of angular scales. Interferometric observations of structure within the resolution limit can implicitly constrain the structure at finer angular scales somewhat beyond it.

Given a set of interferometric data and a compact set of eigenimages, the problem of image reconstruction and model fitting reduces to finding the relative weights of the eigenimages that are necessary for their weighted linear combination to be consistent with the data. It is important to emphasize, however, that while the image space of simulated images is completely sampled by the PCA basis, the EHT coverage provides only sparse, incomplete sampling of the visibility space.  As such, the basis functions in that space (i.e., the visibility maps of the eigenimages) are no longer orthogonal when sampled only at the discrete EHT baselines. As a result, their coefficients must be fitted to the data with a procedure that respects the resultant covariances that now appear when the PCA components are fitted to the visibilities.

The goal of this paper is to progress from the initial presentation of the PCA image reconstruction methodology introduced by \citet{2018ApJ...864....7M} to a complete description of how to apply it to analyzing the EHT observations of accreting supermassive black holes.  In Section \ref{sec:PCAsims}, we describe the GRMHD simulations that we used to construct the PCA basis, the preprocessing of the simulated images, and finally the PCA basis that we derived from them. In Section \ref{sec:MCMC}, we describe the MCMC algorithm we use to fit interferometric data in order to obtain posteriors over the relative weights of the PCA components. We present results of applying \texttt{PRIMO} to simulated interferometric data in Section \ref{sec:results} and summarize our work in Section \ref{sec:discussion}.

\section{Building a PCA Basis From GRMHD simulations}\label{sec:PCAsims}

As outlined in \citet{2018ApJ...864....7M}, we perform PCA on images generated from GRMHD simulations to describe the image space in which EHT images of real accreting black holes are likely to reside. In this section, we detail the methodology used to derive the linear combination of PCA components needed to fit a given data set.

\subsection{The GRMHD Simulations}\label{sec:sims}

The GRMHD simulation images employed to generate the PCA basis were created using the massively parallel GPU-based code {\tt GRay} \citep{2013ApJ...777...13C}. As input to the radiative transfer and ray-tracing simulations, we use two high-resolution GRMHD simulations with long time spans that were created using the 3D {\tt HARM} code \citep{2003ApJ...589..444G, 2012MNRAS.426.3241N, 2013MNRAS.436.3856S}. 

The configuration of a GRMHD simulation is specified by a set of physical parameters.  For the purposes of validating our algorithm, we generated a set of 30 simulation runs, with parameters covering a wide range of possible emission models of the inner accretion flow around the black hole in M87, as follows:

\begin{figure*}[t!]
\centering
\includegraphics[width=1.1\textwidth]{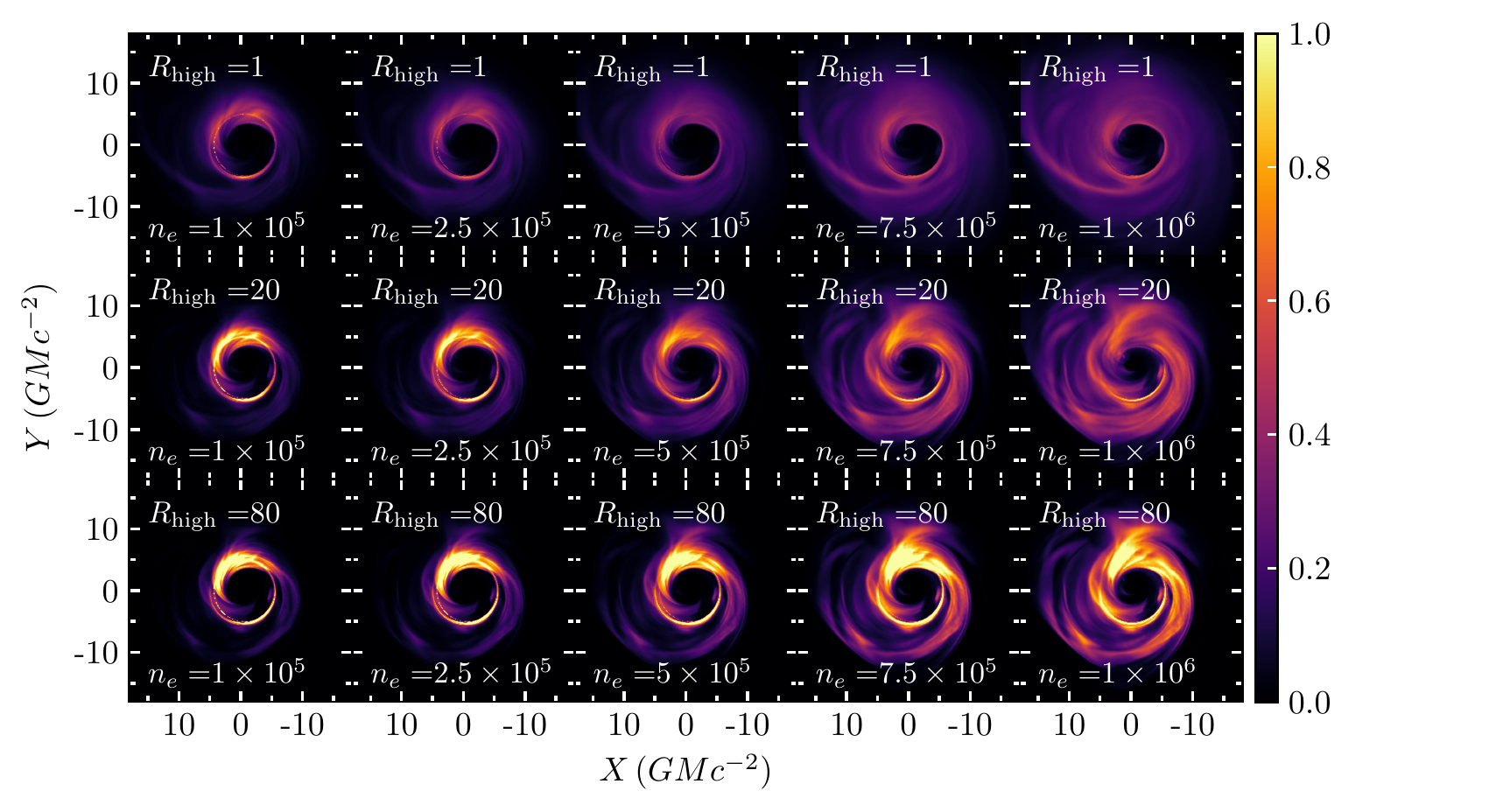}
\caption{Effect of changing the electron density scale $n_e$ (different columns) and the ion-to-electron temperature ratio $R_{\mathrm{high}}$ (different rows) on a single snapshot from a GRMHD simulation.  This snapshot is from a MAD simulation with a black-hole spin of $a=0.9$ pointing upwards in each panel, with an observer inclination of $i=17^\circ$, and a black-hole mass of  $M=6.5\times10^9M_{\odot}$;  the brightness in each panel is normalized such that panels with the same value of $n_e$ have the same total flux. Increasing the electron density scale leads to images with large ring widths whereas changing the temperature ratio alters the relative brightness of the accretion flow and funnel regions.}
\label{fig:ne_rhigh}
\end{figure*}

\begin{itemize}
\item GRMHD simulations only evolve the energy density of the plasma and, therefore, primarily the temperature of the ions and not of the electrons. In the accretion flow, the ion-to-electron temperature ratio is expected to be determined primarily by the plasma $\beta\equiv P_{\mathrm{gas}}/P_{\mathrm{mag}}$ parameter, which is the ratio of the local gas to magnetic pressures~\citep{2015ApJ...799....1C}. In the polar funnel, which is magnetically dominated, the two temperatures are expected to be nearly equal due to magnetic conduction~\citep{2015MNRAS.454.1848R}. In order to capture this behaviour, we used a prescription for the electrons that sets the ion-to-electron temperature ratio $T_{\rm i}/T_{\rm e}$ to \citep{2016A&A...586A..38M, 2019ApJ...875L...4E}
\begin{equation}
\frac{T_i}{T_e}=R_{\mathrm{high}}\frac{\beta^2}{1+\beta^2} +\frac{1}{1+\beta^2}.
\end{equation}
We explore three values for $R_{\mathrm{high}} = 1,\, 20,\, 80$, but note that the $R_{\mathrm{high}}=1$ simulations effectively result in an electron temperature that is equal to the ion temperature throughout the plasma, which is inconsistent with the assumption of a radiatively inefficient flow. We choose to include the $R_{\mathrm{high}}=1$  simulations in our library only for consistency with previous EHT publications and in order to explore a broad, albeit somewhat unphysical, range of image structures.

\begin{figure*}[t!]
\centering
\includegraphics[width=1.1\textwidth]{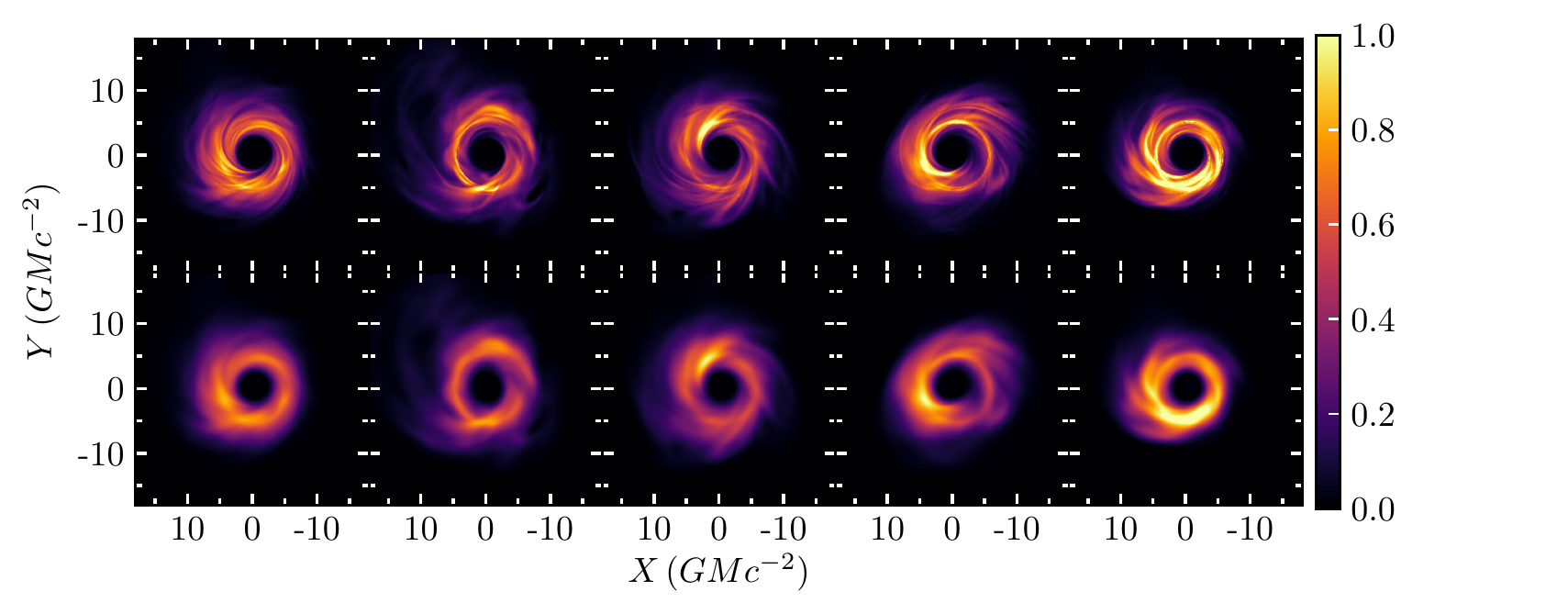}
\caption{(top row) Five representative snapshots from the MAD simulation with $n_e = 5\times10^5$, $R_{\mathrm{high}} = 20$, $i=17^{\circ}$, and $M=6.5\times10^9M_{\odot}$. Even within a single simulation, there is significant structural variability between the various snapshots. 
 (Bottom row) The same snapshots as the top row but filtered using a Butterworth filter with $n=2$ and $r=15 G\lambda$, to mimic the finite resolution of the EHT. The brightness in each panel has been normalized such that they all have the same total flux.}
\label{fig:snapshots}
\end{figure*}

\item The electron density scale provides an overall normalization that sets the total accretion rate in the simulation. We explored values for the electron density scale of $n_e = 10^5,\, 2.5\times 10^5,\, 5\times 10^5,\, 7.5\times 10^5,\, 10^6 \,\mathrm{cm}^{-3}$. We note that the higher values of electron number density are unlikely for M87, given the measured 1.3~mm flux and polarization signatures~\citep{2019ApJ...875L...5E,2021ApJ...910L..13E}, but we include them in our simulation data set for completeness.

\item  In half of the simulations, we used initial conditions that resulted in strong, ordered magnetic fields and a magnetically arrested disk (MAD, see e.g., \citealt{2012MNRAS.426.3241N}); in the other half, we used initial conditions that resulted in a less-ordered, weaker, magnetic field, commonly referred to as standard and normal evolution (SANE, see e.g., \citealt{2003ApJ...592.1042I}).

\item We set the inclination angle of the black hole spin axis relative to the observer's line of sight to $i=17^{\circ}$. This parameter only enters the radiative transfer calculation and determines the relative asymmetry of the image (see, e.g., \citealt{2022ApJ...924...46M}). We made this choice under the assumption that the spin axis of the black hole is parallel to the large scale jet that has been observed at radio wavelengths \citep{2018ApJ...855..128W}. In the PCA model described below, we will allow for the possibility that the spin axis is either aligned or anti-aligned with the large scale jet as well as for an arbitrary position angle of the spin axis in the plane of the sky. Even though the last two considerations affect the orientation of the black-hole image in the sky, they are trivial geometric transformations and do not enter the GRMHD simulations.

\item  We set the black hole mass to $M = 6.5\times10^9 M_{\odot}$ for the initial preparation of the simulations, which is a value consistent with the one obtained by stellar dynamics~\citep{2011ApJ...729..119G} and by the first EHT imaging results~\citep{2019ApJ...875L...6E}. Changing this value has two effects on the resulting simulations. First, it rescales the linear size of each image by a factor proportional to the mass. Second, it affects the outcome of the radiative transfer calculations by altering the synchrotron emission/absorption coefficients and by rescaling the photon path lengths. For the former effect, which is a trivial geometric transformation, we explore different mass values by rescaling the angular size of the PCA basis. For the latter effect, we note that, in the relevant range of parameters, the black-hole mass is nearly degenerate with the electron number density scale $n_e$, with the image brightness at each pixel scaling as $\sim n_{e}^2 M$ (see Appendix A of \citealt{2022ApJ...925...13S}; see also \citealt{2015ApJ...799....1C}). By exploring a broad range of values for the electron density scale and allowing for a rescaling of the images, we effectively probe a broad range of black-hole masses.

\item We assumed a single black hole spin parameter of $a=0.9$ for simplicity since image morphology is only weakly dependent on spin~\citep{2019ApJ...875L...5E}. Indeed, as we show in later sections, the same PCA basis can also be used to reconstruct images of black holes with other spins. 
\end{itemize}

\begin{figure*}[t!]
\centering
\includegraphics[width=1.05\textwidth]{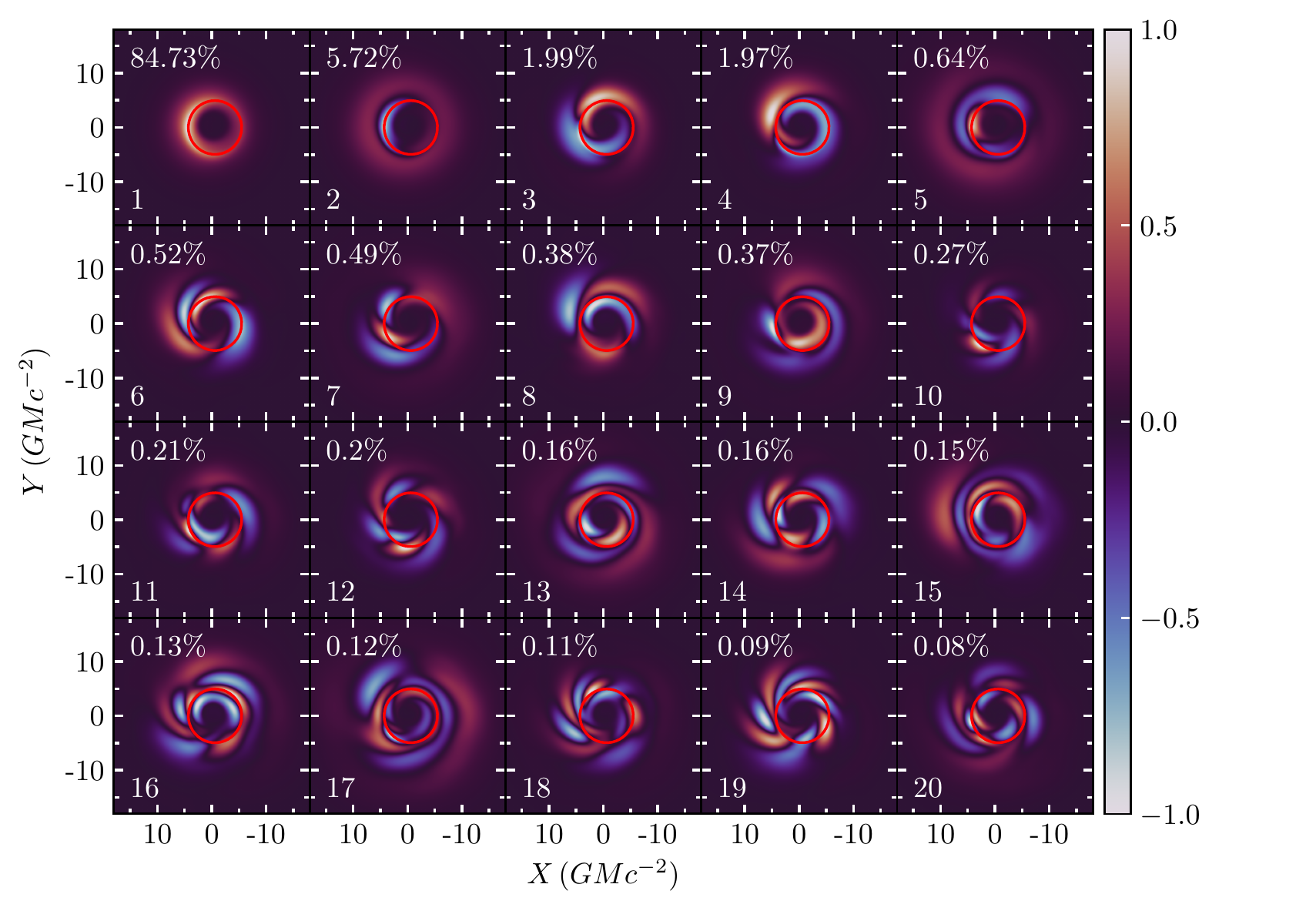}
\caption{The first 20 PCA components for the set of simulated images employed in this paper. The respective proportion of variance explained are shown at the top left corner of each panel. Red closed curves show the analytically calculated size and shape of the black hole shadow boundary, defined as the critical impact parameter between photons that fall into the black hole and those that escape, as seen by an observer at infinity. The first PCA component is similar to the average image of the simulations while contributions from the second component lead to thicker and thinner rings. The third and fourth components allow for either up-down or left-right asymmetries. The higher order components describe smaller scale structure indicative of the variance seen in the simulated snapshots. Each panel is normalized such that the full range of values falls within the color bar range.}
\label{fig:comps_all}
\end{figure*}

Figure \ref{fig:ne_rhigh} shows the effect of changing the electron number density scale, $n_e$, and the ion-to-electron temperature ratio $R_{\mathrm{high}}$ on a single snapshot from a MAD simulation. The electron number density scale affects primarily the width of the bright ring with the latter increasing significantly with increasing $n_e$~\citep{2022ApJ...925...13S}. In contrast, the temperature ratio $R_{\mathrm{high}}$ affects the relative brightness of different parts of the flow, altering the relative brightness between the funnel region and that of the accretion flow.

The set of parameters we discussed reflects a decision as to which sources of image variance to include in the PCA analysis and which parameters to treat externally. The position angle ($\phi$) of the image on the sky, for example, can be included in our model trivially by an overall rotation of the PCA components and need not be included in the derivation of the components themselves. Whether the spin axis is pointing towards us at $17^{\circ}$ or away from us at the complementary angle can also be incorporated in a similar manner, as it describes (statistically) a simple reflection. The effect of the black hole mass on image morphology is mostly degenerate with the electron density except for a change in the overall size of the image, which can be included trivially in the PCA model as a scaling of angular distances applied to all components. The overall source position is also not included in the PCA basis since the current set of EHT data only involve visibility amplitudes and closure phases, which are independent of the image location. 

For each set of parameters, we generated 1024 image snapshots with a time resolution of $10\,GM/c^3$. For the mass of M87, the time resolution equals $\sim 3$ days and 17 hours and each simulation covers a total time span of over ten years. Each snapshot has a field of view of $64 \,GM/c^2$ and a resolution of $1/8\, GMc^{-2}$ per pixel (approximately $0.5\,\mu\mathrm{as}$ resolution). Critically, the field of view is substantially larger than the $\sim 10\,GM/c^2$ measured size of the image and the resolution scale is sufficiently fine to avoid deleterious aliasing effects \citep{2020arXiv200406210P}. 

The set of 30 simulations provides a total of 30,720 images covering a broad range of image morphologies. Figure~\ref{fig:snapshots} shows several snapshots from a single simulation. Here we emphasize that although the parameters of the radiative transfer simulations can significantly affect gross image properties, such as the width of the ring of emission (see Figure \ref{fig:ne_rhigh}), there is significant variance in image morphology even within a single simulation because of the stochastic nature of the MHD turbulence in the accretion flow (see also  \citealt{2017ApJ...844...35M,2018ApJ...856..163M,2018ApJ...864....7M}).

\subsection{Preparing the Simulated Images for PCA}

The simulated images have significant structure at small scales, which the EHT cannot probe. Because we want the PCA basis to only reflect image variance on the physical scales observed by the EHT, we first need to eliminate the high spatial-frequency structure in each simulated image.

To achieve this, we use a Butterworth filter \citep{butterworth1930}, which is effectively a low-pass filter, having a flat response for low Fourier frequencies and declining to zero smoothly at high-frequencies. The Butterworth filter is defined as
\begin{equation}
    F_{\mathrm{BW}}(b) = \left[ 1+\left(\frac{b}{r}\right)^{2n} \right]^{-1/2},
\end{equation}
where $r$ is the scale of the filter and $n$ is a power-law index. We discuss in detail the motivation for using a Butterworth filter as well as the choice of parameters for EHT data analysis in \citet{2020arXiv200406210P}. The bottom row of Figure \ref{fig:snapshots} shows the snapshots of the top row filtered by a Butterworth filter with $n=2$ and $r=15G\lambda$. This choice of filter parameters allows us to retain most of the power at baseline lengths probed by the EHT array, while filtering out most of the power at larger lengths.

As a second step, we normalize each filtered image to have the same total flux. Because images with higher electron density scale $n_e$ have significantly higher total flux, not normalizing would have biased the PCA basis towards images with higher $n_e$ values. We explored the effects of standardizing the images by their variance and found that this has a negligible effect on the PCA basis other than on the overall normalizations. We, therefore, do not standardize the images by their variance. We also do not mean subtract the images before performing PCA, i.e., similar to what was done in \citet{2018ApJ...864....7M}, since the properties of the mean image are critical in fitting the observed data. If, instead, we had mean subtracted the images before performing PCA, we would have needed to add back the mean image to the linear combination of PCA components, resulting in the same number of free parameters in the model. 

Since all of the images correspond to the same black hole spin $a=0.9$ and the same inclination angle $i=17^{\circ}$, all of the black hole shadows are concentric and aligned with each other. For the case of M87, this is justified because of the known inclination of the large-scale jet as well as the weak dependence of the simulated images on black-hole spin. If that were not the case, we would have also needed to recenter and align the images before performing PCA, along the line of the approach in~\citet{2020ApJ...896....7M}. 

\subsection{Building the PCA Basis}\label{sec:PCA}

\begin{figure}[t!]
\centering
\includegraphics[width=\columnwidth]{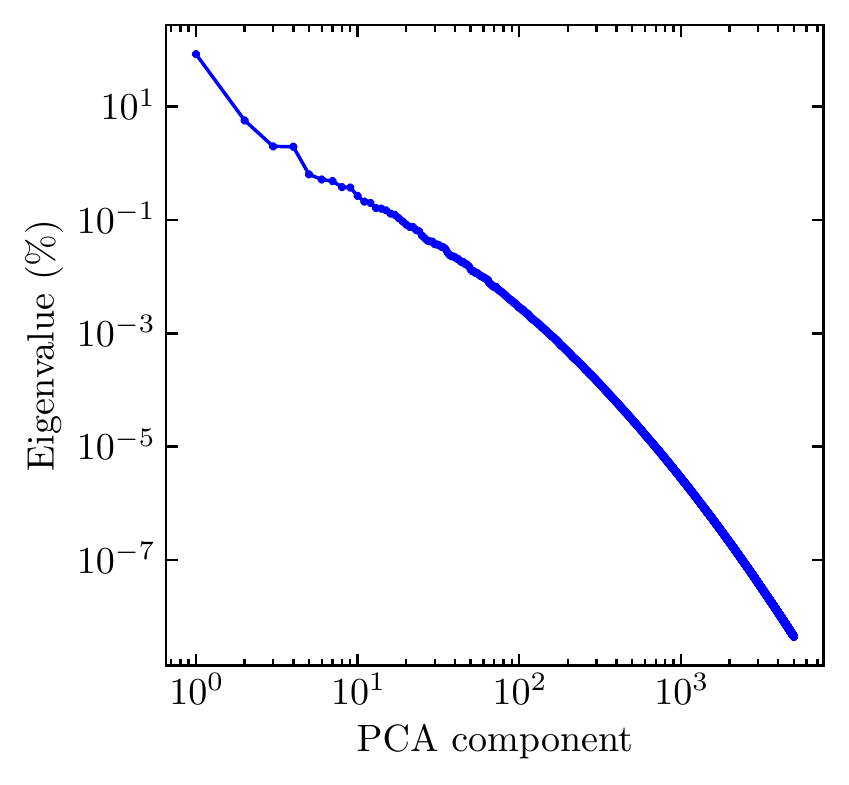}
\caption{The spectrum of normalized eigenvalues for the first 5000 PCA components derived from a set of 30,720 GRMHD snapshots. Only ten PCA components are necessary to reconstruct $98\%$ of the total variance, whereas twenty PCA components can recover $99\%$ of it.}
\label{fig:vals_all}
\end{figure}

Given the complete set of filtered simulated images, we generated the PCA basis following the procedures established in \citet{2018ApJ...864....7M}.
Figure \ref{fig:comps_all} shows the first 20 PCA components. The first PCA component is similar to the average image and contains a positive flux. The higher order PCA components contain both positive and negative fluxes, since these components re-distribute the flux present in the first component to approximate each individual snapshot. 
\begin{figure*}[t!]
\centering
\includegraphics[width=1.05\textwidth]{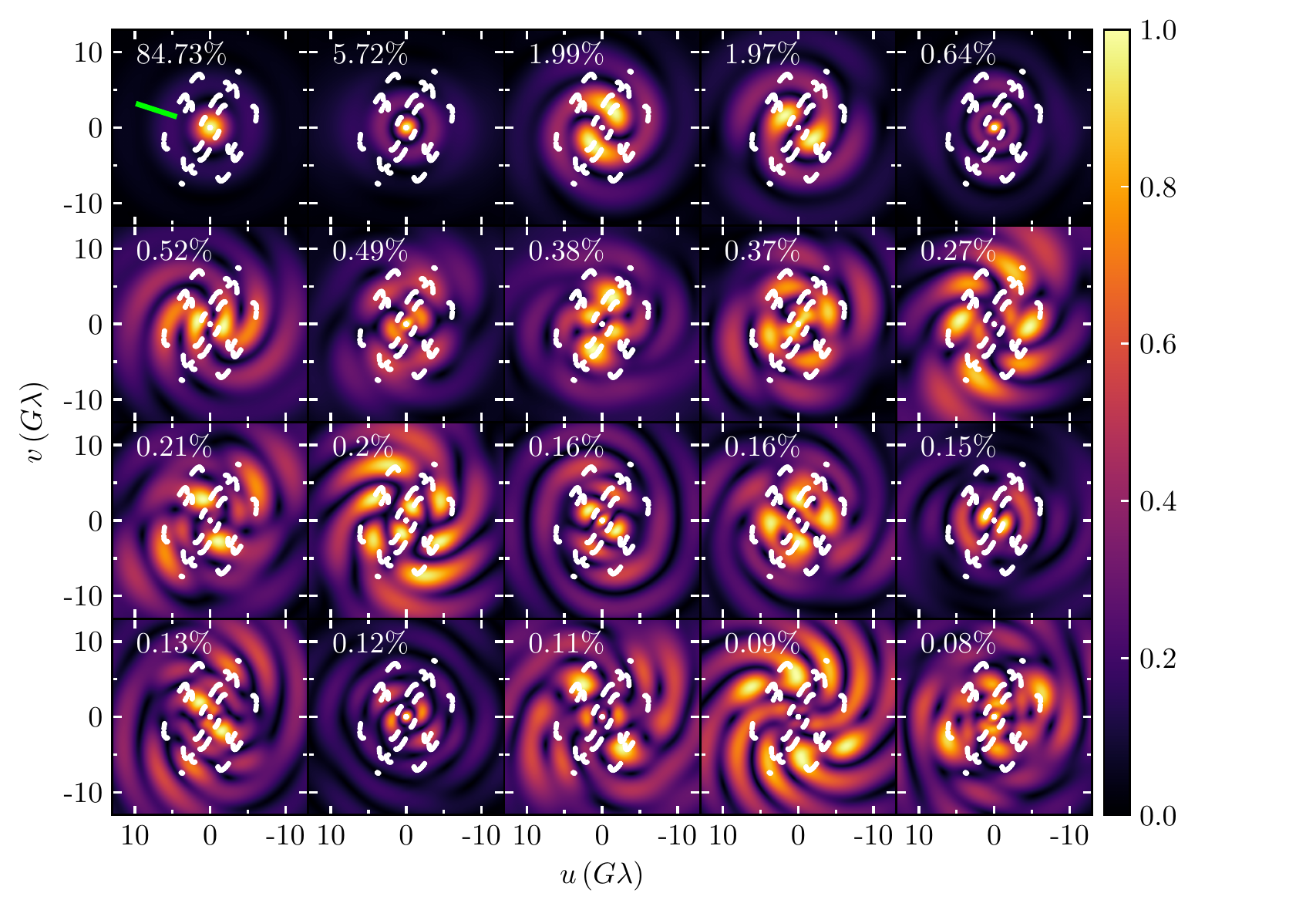}
\caption{The visibility amplitude maps of the first 20 PCA components shown in Figure \ref{fig:comps_all}, normalized such that the peak in each component is equal to unity. The peak values in visibility amplitude for many of the components are offset from the center, indicating that there is significant negative flux in the images, as can be seen in Figure~\ref{fig:comps_all}. White curves show the baseline tracks during the 2017 April EHT observations for the black hole in M87. In all panels, the black hole spin points upwards and the orientation of North is shown by the green line segment in the upper-left panel.}
\label{fig:comps_VA}
\end{figure*}

The normalized eigenvalues corresponding to each PCA component are shown in the top left corner of each panel. Each eigenvalue measures the variance in pixel brightness of each PCA component, normalized such that the sum of all eigenvalues is equal to unity. Figure~\ref{fig:vals_all} shows the eigenvalue spectrum for this PCA decomposition. The first few PCA components account for the majority of brightness variance in the image and only 20 components are needed to account for 99\% of the variance found in the full set of simulations. The slope of the eigenvalue spectrum for higher components is set by the power spectrum of the structures in the images \citep{2018ApJ...864....7M}. 

\begin{figure*}[t!]
\centering
\includegraphics[width=1.05\textwidth]{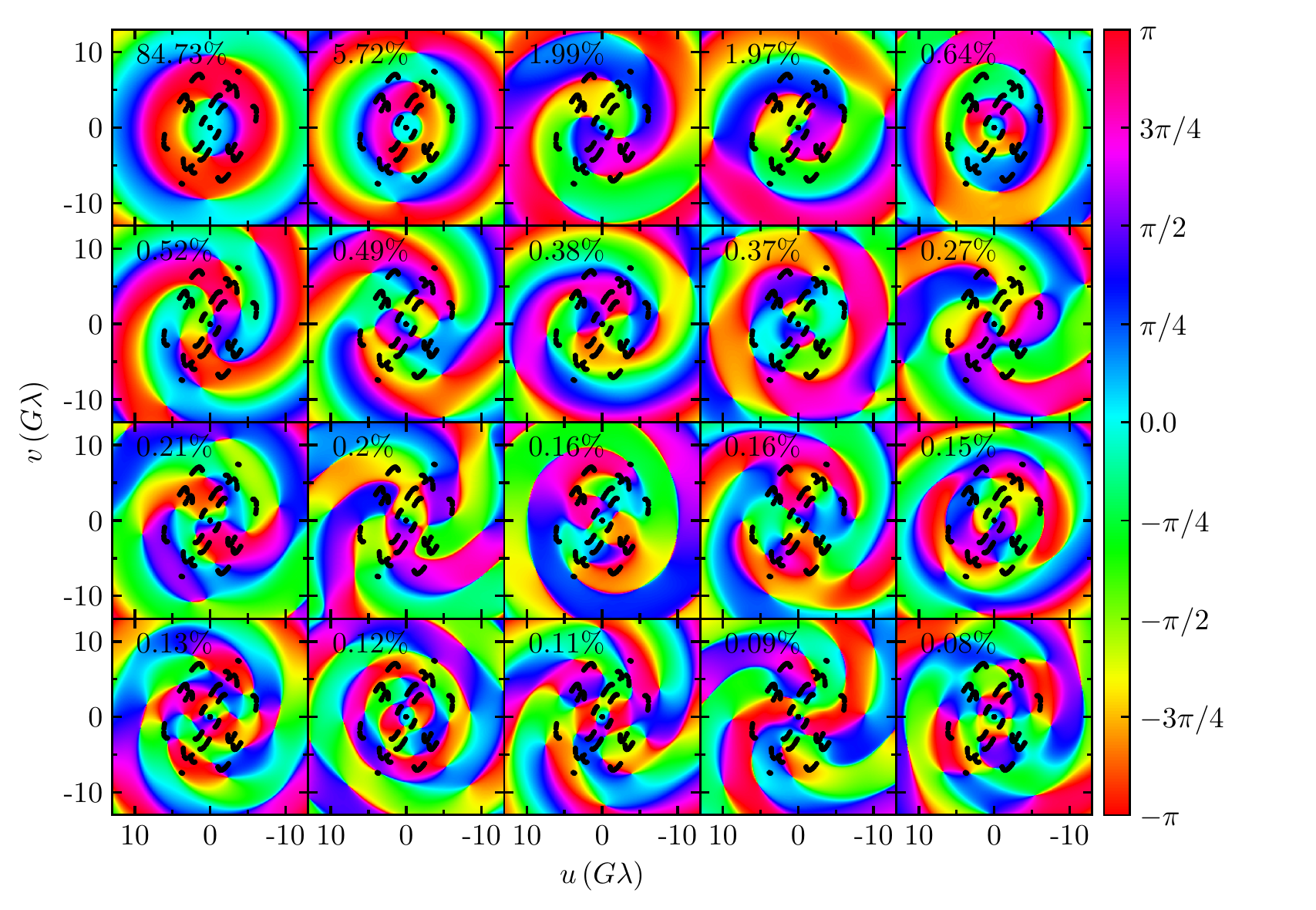}
\caption{Same as Figure~\ref{fig:comps_VA} but for visibility phase maps.}
\label{fig:comps_VP}
\end{figure*}

Figures~\ref{fig:comps_VA} and \ref{fig:comps_VP} show the corresponding visibility amplitude and phase maps of the first 20 PCA components. It is a linear combination of these components in visibility space that we will fit directly to the data. As expected, the first few components contain primarily structures with low spatial frequencies (i.e., small baseline lengths) and describe primarily the broad-brush structure of the image. The remaining components contain significant power at high spatial frequencies (i.e., large baseline lengths) and describe the smaller structures in the image. 

It is interesting that, although this was not explicitly imposed when performing the principal component decomposition, components of increasing PCA order correspond to higher order ($m$-fold) azimuthal symmetry. This is important when comparing the angular structure of the PCA components to the locations of the EHT baselines for the 2017 M87 observations~\citep{2019ApJ...875L...3E}, as also shown in Figures~\ref{fig:comps_VA} and \ref{fig:comps_VP}. Note that we have rotated the baseline tracks such that the black-hole spin axis, which points upwards in all these panels, is at $288^{\circ}$ East of North. Clearly, the first 20 PCA components already incorporate a substantial degree of azimuthal structure, which is finer than the angular separation of the dominant locations in visibility space probed by the EHT array. Lastly, note that each component comprises detail over a broad range of spatial frequencies.  Within a given component, structural information on fine angular scales is correlated with that on broader scales.  This allows visibilities within the EHT band limit to lead to inferences on the structure somewhat beyond it, producing reconstructions with a degree of ``super-resolution.''

\smallskip

\section{MCMC algorithm}\label{sec:MCMC}

In order to fit EHT data, we implement the linear PCA model (\texttt{PRIMO}) into the MCMC algorithm MARkov Chains for Horizons (\texttt{MARCH}, \citealt{2022ApJ...928...55P}). For the purposes of this initial exploration, we fit this model to synthetic EHT data calculated for the baseline tracks of the array during the 2017 April 5th observations of M87.

\subsection{The PCA Image Model}

The PCA decomposition described in Section 2 allows us to construct a model for a black-hole image that is a linear combination of $N$ PCA components, with an appropriate rescaling, to account for a different black-hole mass, and an appropriate rotation, to allow for different orientations in the sky.

We will be fitting data in the visibility domain and, therefore, define the linear combination of the first $N$ PCA components in that domain as
\begin{equation}
   \tilde{I}(u,v) = \sum_{n=1}^{N} a_n {\bf \tilde{u}}_n(u,v), 
\end{equation}
where ${\bf \tilde{u}}_n$ are the PCA components in the Fourier ($u,v$) domain, $\tilde{I}$ is the Fourier domain visibility of the reconstructed image, and $a_n$ is the amplitude of the $n-$th PCA component. Without loss of generality and in order to facilitate comparison with other astrophysical measurements of the sources, we set $a_1=1$ and instead fit for the total zero baseline visibility amplitude, which is also equal to the image flux
\begin{equation}
F=\sum_{n=1}^N a_n {\bf \tilde{u}}_n(0,0)\;.
\end{equation}

By construction, this same linear combination of the PCA components in the image domain also generates the ``best-fit'' image, i.e.,
\begin{equation}
   I({\rm X,Y}) = \sum_{n=1}^{N} a_n {\bf {u}}_n({\rm X,Y}), 
\end{equation}
where now $I$ is the reconstructed image and ${\bf u}_n$ are the PCA components, both in the image domain $(X,Y)$.

In addition to the $N-1$ PCA amplitudes and the flux normalization $F$, the  model also includes three parameters that are implemented as a scaling, a rotation, and an up-down flip of the image.
In particular, we introduce
\begin{itemize}
    \item A scaling parameter $\theta_g={GM}/(Dc^2)$ that is applied to all PCA components in the sky domain (or equivalently $\theta_g^{-1}$ that is applied in the visibility domain). This scaling parameter quantifies the mass-to-distance ratio of the particular black hole we are modeling and allows us to convert the length scales in our images, which are in gravitational units, to angular sizes in the sky. This parameter can also be informed by the strong priors obtained by modeling the dynamics of stars around the black hole~\citep{2019ApJ...875L...6E}.

\item  A position angle $\phi$, measured in degrees East of North, applied to all PCA components, that quantifies the orientation of the black-hole spin on the plane of the sky. 

\item A flip parameter $j=-1,1$ that accounts for the possibility that the spin axis is pointing away from the observer and therefore that the accretion flow is rotating in a sense that is opposite (i.e., clockwise) to that of the simulation. In other words, if $j=-1$, we mirror all PCA components along the x-axis such that the rotation patterns will be orientated in the clockwise direction. 
\end{itemize}

We note that, for computational efficiency, we do not use the three parameters $\theta_g$, $\phi$, and $j$ to scale, rotate, and flip each of the PCA components. Instead, we use them to scale, rotate, and flip appropriately the small number of discrete $u-v$ locations of the EHT baselines. We then calculate the linear sum of the PCA components in these locations using the interpolation technique we discuss below.

In total, the PCA model has $N+3$ free parameters, where $N$ is the number of PCA components used. Finally, it is worth emphasizing that, even though the PCA model is linear in most of its parameters, the visibility amplitudes and closure phases that we fit it to involve non-linear operations.

\subsection{Two-Dimensional Interpolation}

At each step of the MCMC chain, the algorithm calculates the model prediction at the $(u,v)$ location of each data point and compares it to the data. Since the PCA image model is numerical and sampled on a regular array of pixels, we evaluate its prediction at any desired location using a 2D sinc interpolation, which has been demonstrated to cause no degradation of resolution of the 2D maps~\citep{1986ftia.book.....B}. In 1-D, a sinc function is defined as
\begin{equation}
    \mathrm{sinc}(u) = \frac{\sin(\pi u)}{\pi u},
\end{equation}
where $u$ is the pixel coordinate in the Fourier domain.  Interpolation in 2D is done with separable sinc kernels in $u$ and $v$ that are multiplied to form a 2-D kernel.  

Along each orientation, the value of the visibility at $u'$ is given by
\begin{equation}
    f(u') = \sum_{n} \frac{\sin[\pi (n-\Delta u)]}{\pi (n-\Delta u)}f(n), 
\end{equation}
where $f(n)$ is the image value at the integer $n$ locations, and $\Delta u= u'-u.$ In practice, we limit the kernel to a finite domain of $\pm u_0,$ and taper it smoothly with a Gaussian to produce a well-behaved cutoff in the Fourier domain, 
\begin{equation}
    f(u') = \frac{1}{C_{\mathrm{sinc}}}\sum^{u_0/\Delta u}_{n=-u_0/\Delta u}e^{-(n-\Delta u)^2/2\sigma^2}~ \frac{\sin[\pi (n-\Delta u)]}{\pi(n-\Delta u)}f(n),
\end{equation}
where $\sigma$ is chosen such that 2-3 cycles of the sinc function are included. The normalization constant $C_{\mathrm{sinc}}$ ensures that the interpolation kernel has an integral of unity, given the tapering and finite domain: 
\begin{equation}
    C_{\mathrm{sinc}} = \sum^{u_0/\Delta u}_{n=-u_0/\Delta u}e^{-(n-\Delta u)^2/2\sigma^2}~\frac{\sin[\pi (n-\Delta u)]}{\pi(n-\Delta u)}.
\end{equation}
However, since the $\sin[\pi(n-\Delta u)]$ term is periodic with an amplitude specified by the $\Delta u$ phase, its particular value, but for an alternating sign, is constant and thus is absorbed in the normalization.  In practice, evaluating a trigonometric function is not required, since we can write  
\begin{equation}
    f(u') = \frac{1}{C_{\mathrm{sinc}}'} \sum^{u_0/\Delta u}_{n=-u_0/\Delta u} \frac{e^{-(n-\Delta u)^2/2\sigma^2} (-1)^n}{(n-\Delta u) }f(u)
\end{equation}
where
\begin{equation}
    C_{\mathrm{sinc}}' = \sum^{u_0/\Delta u}_{n=-u_0/\Delta u}\frac{e^{-(n-\Delta u)^2/2\sigma^2}(-1)^n}{(n-\Delta u)}.
\end{equation}

\subsection{The Posterior Distribution}\label{sec:posterior}

Having defined a visibility-domain PCA model that depends on $N+3$ model parameters, which we collectively denote by the vector $\vec{\theta}$, we use Bayes' theorem to write the posterior over these parameters as 
\begin{equation}
P(\vec{\theta}\vert{\rm data})=C\; P_{\rm pri}(\vec{\theta})\; {\cal L}({\rm data}|\vec{\theta})\;.
\label{eq:bayes}
\end{equation}
Here, $P_{\rm pri}(\vec{\theta})$ is the prior distribution over the model parameters, ${\cal L}({\rm data}|\vec{\theta})$ is the likelihood that the set of observations can be obtained from the model, and $C$ is an appropriately defined normalization constant.

The set of data obtained by the EHT is a series of visibility amplitudes at the various baseline lengths between the different pairs of stations as well as a series of closure phases along all possible baseline triangles~\citep{2019ApJ...875L...3E}.  We calculate the likelihood function by multiplying the likelihoods of the individual visibility amplitude and closure phase data (see, however,~\citealt{2019ApJ...882...23B}), assuming that all likelihoods are independent of each other
\begin{equation}
{\cal L}({\rm data}|\vec{\theta})= \prod_i {\cal L}_i({\rm data}|\vec{\theta})\;.
\label{eq:like}
\end{equation}
The precise definition of the various likelihoods is provided in detail in \citet{2022ApJ...928...55P}. Because they depend only on the data products, they are the same for all models. The priors over the model parameters, however, are specific to each model, as we discuss in detail in the following subsection.

\subsection{Priors}\label{sec:Priors}

To ensure that our PCA model is probing physically relevant areas of the parameter space, we include a combination of informative and non-informative priors on the various model parameters. 

Because the EHT is an interferometer, the total flux $F$ of the compact image cannot be directly measured without perfect knowledge of the prior calibration of the various telescope gains. However, it can often be independently constrained using other single-dish observations. Due to extended emission, the zero-baseline flux of the M87 EHT data was significantly higher than what was reasonably expected for the compact source (see the discussion in appendix B of \citealt{2019ApJ...875L...4E}). Because of this, most EHT M87 analyses constrained the zero-baseline flux to a well-motivated value. To mirror those analyses, we fix the zero-baseline flux at 0.6 Jy, the value used to generate the synthetic data.


\begin{figure*}[t!]
\centering
\includegraphics[width=\textwidth]{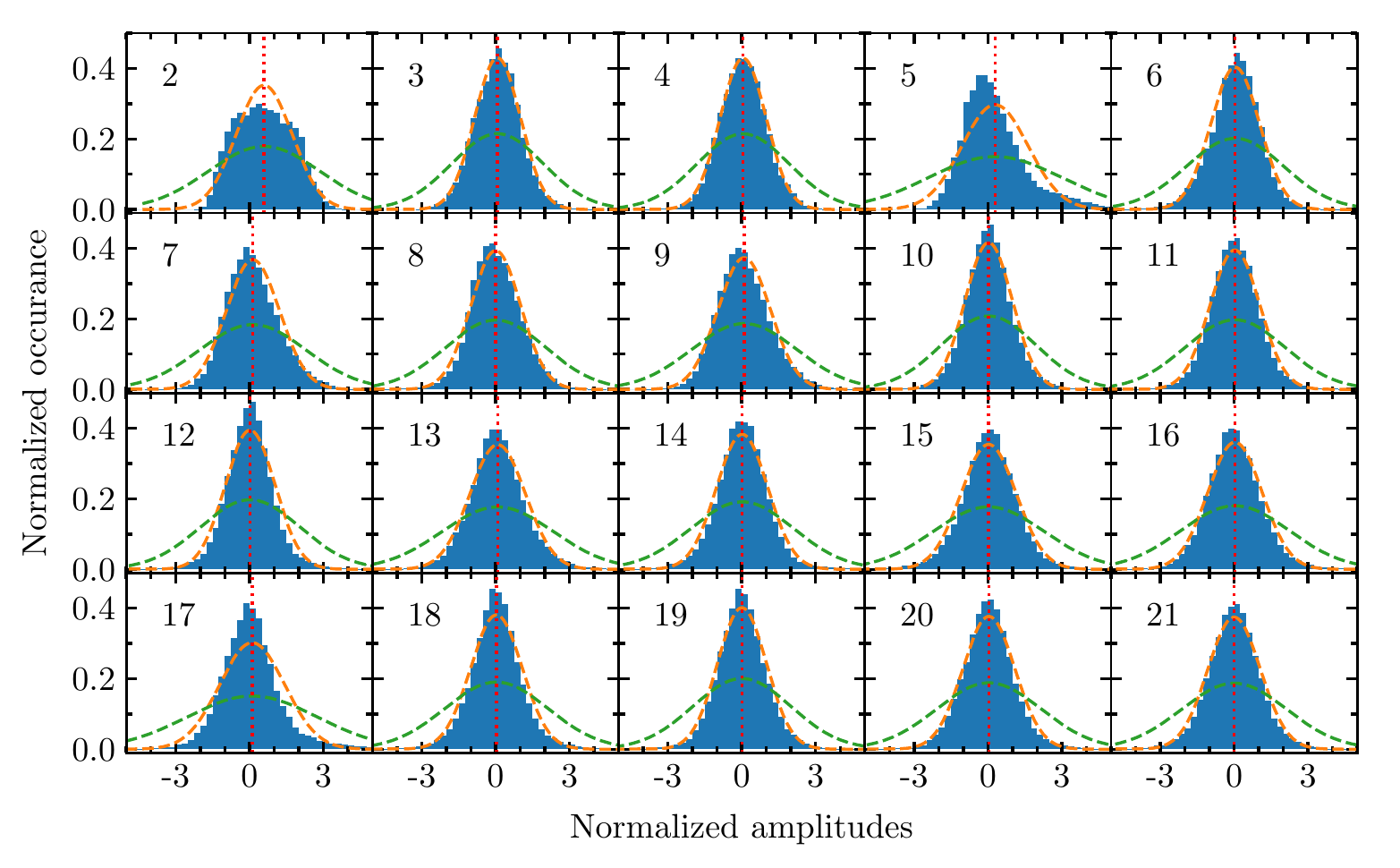}
\caption{The distribution of the normalized amplitudes, $a_n/a_1$, for PCA components 2 through 21. The red dotted lines show the mean of the distribution in each panel and the dashed orange lines show Gaussians with widths set by the standard deviations of the distributions and with peaks at the means of the distributions. The green dashed curves show the Gaussians broadened by a factor of two. We use the broadened Gaussians as priors to allow for reconstruction of images that are outliers within the distributions as well as of images that are similar to but are not contained in our training set.}
\label{fig:amps_grid}
\end{figure*}

For the scaling parameter $\theta_g$, there often exist  prior measurements based on gas and/or stellar dynamics. For the M87 black hole, the two measurements are not statistically consistent with each other (see \citealt{2011ApJ...729..119G, 2013ApJ...770...86W}). The envelope of the credible intervals for these two measurements is contained within the conservative range $1~\mu$as$~\le \theta_{\rm g}\le~6~\mu$as. For this reason, we simply use an uninformative prior
\begin{equation}
  P({\theta_g})=\left\{\begin{array}{ll} \theta_g^{-1}\,\,\,\,\, \mathrm{if}\,1\,\mu\mathrm{as}\le\theta_g\le 6\,\mu\mathrm{as}\\ 0\,\,\,\,\, \mathrm{otherwise.}\\ \end{array} \right.  
\end{equation}

For the orientation parameter $\phi$, we employ a highly informative prior based on the assumption that the black-hole spin is either aligned or anti-aligned with the large-scale jet observed at longer wavelengths, i.e., that
\begin{equation}
    P(\phi) = \frac{1}{2\sqrt{2\pi\sigma_{\phi}^2}}\left[ e^{-(\phi-\phi_0)^2/2\sigma_{\phi}^2} +e^{-(\phi-\phi_0+\pi)^2/2\sigma_{\phi}^2} \right].
\end{equation}
Here $\phi_0=288^\circ$ is the orientation of the large scale jet \citep{2018ApJ...855..128W}. We set the widths of the two Gaussians to a nominal value of $\sigma_{\phi}={\pi}/{8}$. We allow the flip parameter $j$ to be equal to either 1 or -1, with the same prior. 

Finally, we employ informative priors on the amplitudes of the PCA components. Our aim is to give higher priors to images for which the amplitudes of the PCA components are not very dissimilar from the amplitudes that correspond to the simulated images used to calculate the PCA decomposition. However, we also do not wish to limit the fit to images that have precisely the same range of amplitudes as the training set. To achieve this, we first calculate the distribution of amplitudes for each PCA component found in the ensemble of training images and then broaden this distribution by a factor of two. 

Figure~\ref{fig:amps_grid} shows the distribution of normalized amplitudes, $a_n/a_1$, for the PCA components 2 through 21 that we calculated above; note that, by definition, we have set $a_1=1$. Each panel also shows a Gaussian (in orange) with the same mean and standard deviation as the numerical distribution. These Gaussians provide good descriptions of the distributions for almost all of the components shown in the figure, with components two and five being notable exceptions. Both of these components contain structure that controls the width of the ring in the image, which is strongly dependent on the simulation parameters (e.g., $n_e$). Therefore, the distributions of amplitudes for these components are not expected to follow a Gaussian distribution but rather will depend on the particular set of parameters used for the simulation library.

Gaussian distributions with the same mean but twice the standard deviation are also shown in each panel (green dashed lines) and comfortably include the full range of amplitudes found in the training image set.  In practice, for computational efficiency, we use these broadened Gaussians as priors on the amplitudes of each PCA component. In other words, we write the prior for the normalized amplitude of the $n-$th PCA component $a_n/a_1$ as
\begin{equation}
P(a_n/a_1) = \left[e^{-\frac{1}{2}\left( \frac{a_n/a_1-\overline{a_n/a_1}}{2\sigma_n}\right)^2}\right]/({\sqrt{2\pi}2\sigma_n})\;,  
\end{equation}
where $\overline{a_n/a_1}$ and $\sigma_n$ is the mean value and standard deviation of the distribution of normalized amplitudes of the training set.

\begin{figure*}[t!]
\centering
\includegraphics[width=\textwidth]{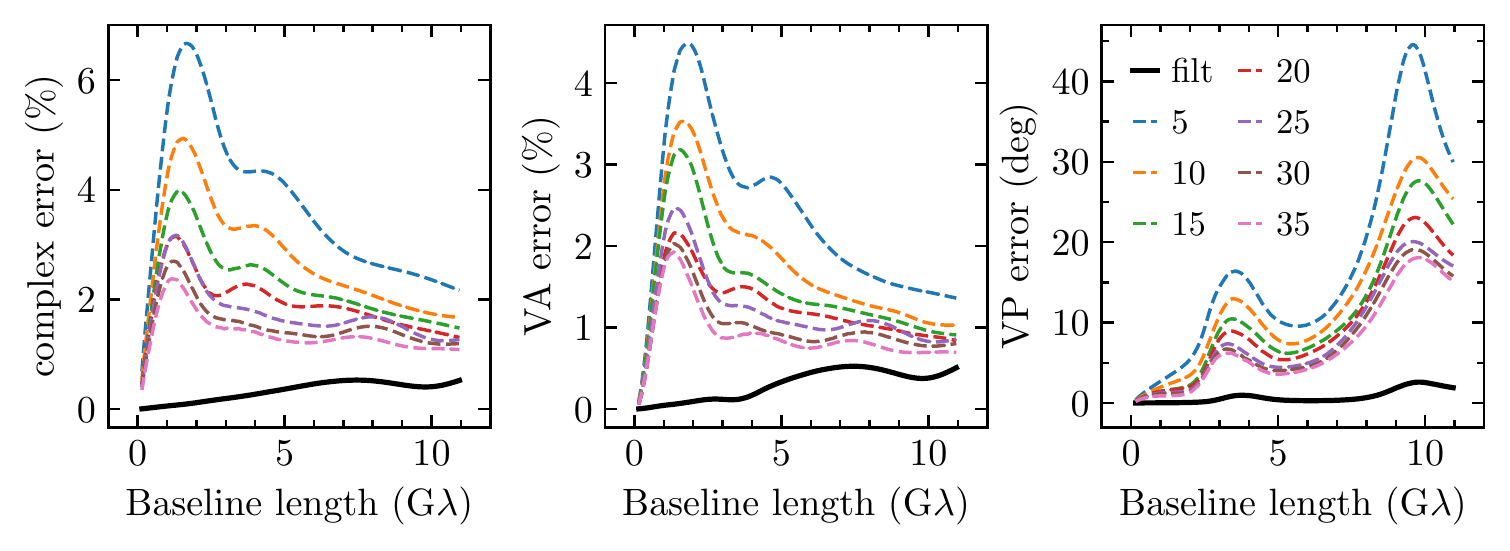}
\caption{Fractional complex error ($\epsilon_{\mathrm{complex}}$, left), fractional error in visibility amplitude ($\epsilon_{\mathrm{VA}}$, middle), and error in visibility phase ($\epsilon_{\mathrm{VP}}$, right) for reconstructions with 5, 10, 15, 20, 25, 30, and 35 PCA components. The black curve shows the error between the original unfiltered snapshot and the snapshot filtered with the Butterworth filter with $r=15\,G\lambda$ and $n=2$. All reconstructions are compared to the original unfiltered snapshot. We calculate these error quantities for each of the 30,720 snapshots, and then average them as a function of baseline length. The longest baseline that the 2017 EHT array could observe was $\sim 8\,G\lambda$. Reconstructions with 20 components achieve errors less than $\sim 3\%$ for $\epsilon _{\mathrm{complex}}$, $\sim 2\%$ for $\epsilon_{\mathrm{VA}}$, and $\sim 15^{\circ}$ for $\epsilon_{\mathrm{VP}}$ for baseline lengths observable by the 2017 EHT array.
\\}
\label{fig:theory_err}
\end{figure*}
\subsection{Theoretical uncertainty}\label{sec:theory_err}

In most applications of PCA, one can reconstruct an image by simply projecting the image onto the PCA components to find the relative amplitude of each component that will result in the best possible reconstruction. Using a higher number of components will invariably result in a higher-fidelity reconstruction. A loss-less reconstruction can always be achieved using all of the PCA components, if the image is part of the original set that was used to calculate the PCA decomposition. In the present application, however, we do not have a full image onto which we can project the components; we instead have sparse $u-v$ coverage. Attempting to fit a large number of components to sparse interferometric data can result in overfitting since there may be several possible linear combinations of components that fit the data. Therefore, there exists an optimal number of PCA components for which the highest-fidelity reconstruction can be achieved by fitting the sparse interferometric data while respecting the resolution of the array.

In order to determine this optimal number and asses the error introduced by the truncation, we quantify the error in the visibility amplitudes between a reconstruction with $N$ components and the original, unfiltered image in the Fourier domain as 
\begin{equation}
    \epsilon_{\mathrm{complex}}= \frac{\sqrt{|(V_{\mathrm{0}}-V_{N})(V^*_{0}-V^*_{N})|}}{F}\;,
\end{equation}
where $F$ is the total flux of the image, $V_{N}$ are the complex visibilities of the reconstruction, vertical bars indicate magnitude, and the asterisk denotes complex conjugation. We define the fractional error in visibility amplitude as
\begin{equation}
\epsilon_{\mathrm{VA}} =\left| \frac{|V_{\mathrm{orig}}| - |V_{\mathrm{recon}}| }{F}\right|
\end{equation}
where $|V_{\mathrm{orig}}|$, and $|V_{\mathrm{recon}}|$ denote the amplitude of the complex visibilities for the original and reconstructed images respectively. The error in visibility phase is defined as
\begin{equation}
    \epsilon_{\mathrm{VP}} =|\arg(V_{\mathrm{orig}}) - \arg(V_{\mathrm{recon}})|
\end{equation}
if this quantity is $<180^{\circ}$ and 
\begin{equation}
    \epsilon_{\mathrm{VP}} =360^{\circ} - |\arg(V_{\mathrm{orig}}) - \arg(V_{\mathrm{recon}})|
\end{equation}
otherwise. We calculate these errors for each baseline length by averaging along different azimuthal orientations and over the complete set of images in the training set.

In both equations above, $\arg(V)$ denotes the argument or phase of the complex visibilities of the images. When taking the average of the error in visibility phase, we follow \citet{mardia2009directional} and define the average of a directional quantity as
\begin{equation}
    \bar{\theta} =
    \begin{cases}
    \tan^{-1}(\bar{S}/\bar{C}), & \mathrm{if\,} \bar{C}\geq 0\\
    \tan^{-1}(\bar{S}/\bar{C})+\pi, & \mathrm{if\,} \bar{C}< 0,
    \end{cases}
\end{equation}
where 
\begin{align}
    \bar{S} &= \frac{1}{n}\sum^{n}_{j=1} \sin{\theta_j}\\
    \bar{C} &= \frac{1}{n}\sum^{n}_{j=1} \cos{\theta_j}.
\end{align}

\begin{figure*}[t!]
\centering
\includegraphics[width=\textwidth]{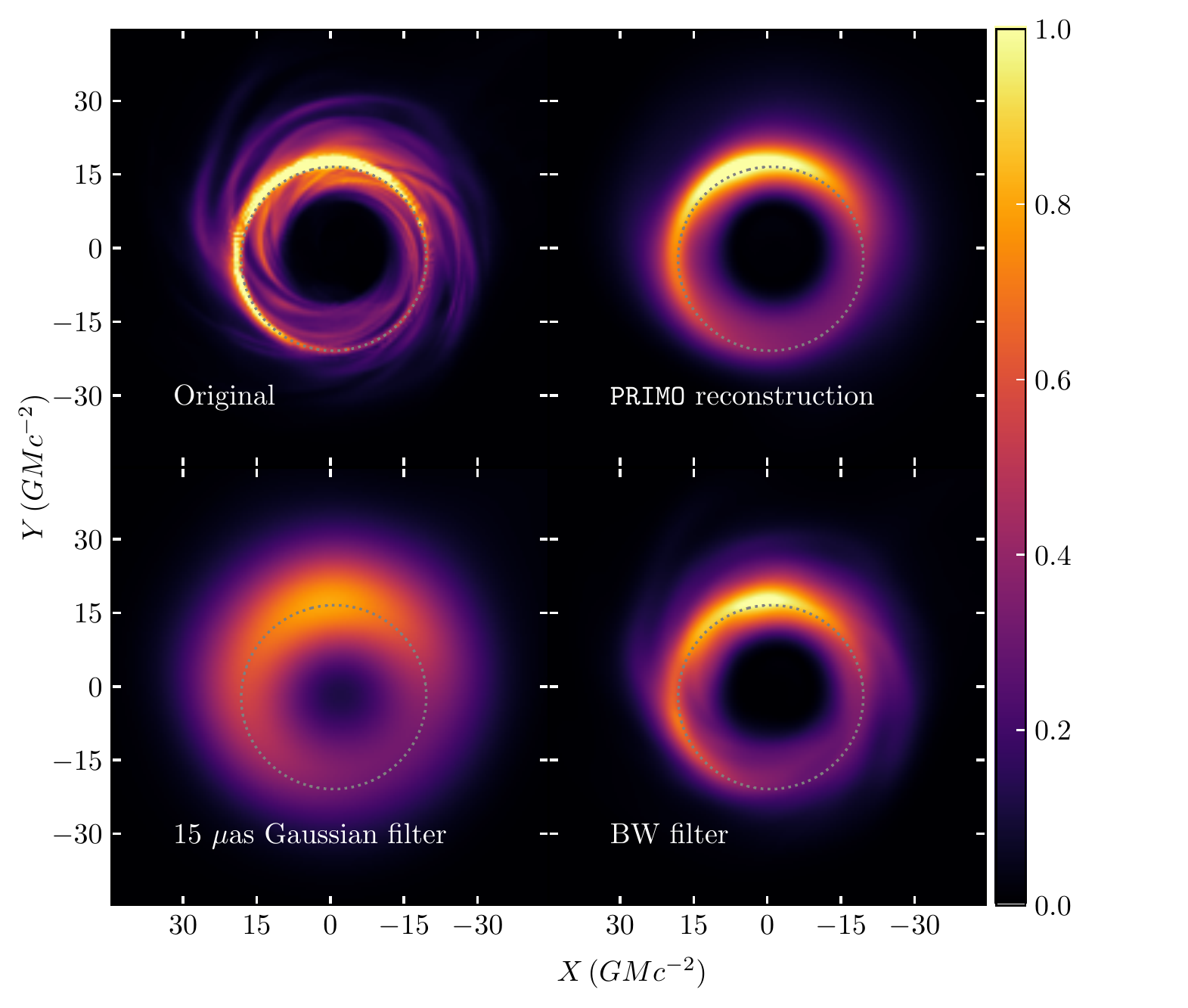}
\caption{{\it (Top left)} Simulated image used for the generation of the first synthetic data set.  {\it (Top right)} The highest likelihood \texttt{PRIMO} reconstruction for the synthetic data. \texttt{PRIMO} can accurately reproduce the depth and shape of the brightness depression, the size and width of the ring of emission, and the brightness asymmetry of the ring. {\it (Bottom left)} The original image blurred by a Gaussian filter with a width of $15\,\mu\mathrm{as}$, which mimics the nominal resolution of the EHT (see text). {\it (Bottom right)} The simulated image convolved with a Butterworth filter with radius $r=15\,\mathrm{G}\lambda$ and index $n=2$. The brightness in all panels has been normalized such that all images have the same total flux, with the exception of the Gaussian broadened image, which has a total flux that is 1.5 times higher than the other panels. In all panels, the gray dotted circles indicate the analytically calculated size and shape of the black hole shadow. The original and filtered images have been rotated to the position angle used to generate the synthetic data set and the \texttt{PRIMO} reconstruction has been rotated by the position angle $\phi$ derived from the model.}
\label{fig:M47}
\end{figure*}

Figure~\ref{fig:theory_err} shows the errors $\epsilon_{\mathrm{complex}}$, $\epsilon_{\mathrm{VA}}$, and $\epsilon_{\mathrm{VP}}$ as a function of baseline length, for all 30,720 snapshots and for different values of the number $N$ of PCA components. The Figure also compares these errors to those introduced to the original images by the application of the Butterworth filter. In all three error quantities, there are significant broad peaks at around $1-4\,G\lambda$, which are introduced by the  dips, or nulls, that exist in the training set around these baseline lengths (see~\citealt{2017ApJ...844...35M} for a discussion of the origin of these uncertainties).

The longest baselines included in the 2017 EHT array are about $8\,G\lambda$. Reconstructions with 20 components achieve fractional complex errors less than $\sim 3\%$ at all baselines less than $8\,G\lambda$, even at baseline lengths that frequently have a significant dip in visibility amplitude. The same reconstructions achieve a fractional error in visibility amplitude of less than $\sim 2\%$ and an error in visibility phase less than $\sim 15^{\circ}$ at all baselines less than $8\,G\lambda$. At baselines that do not coincide with the visibility amplitude minima, the errors are significantly smaller; fractional complex error in visibility amplitude for reconstructions with just 20 PCA components is $\sim 2\%$ in regions between visibility amplitude minima. 

\begin{figure*}[th!]
\centering
\includegraphics[width=\columnwidth]{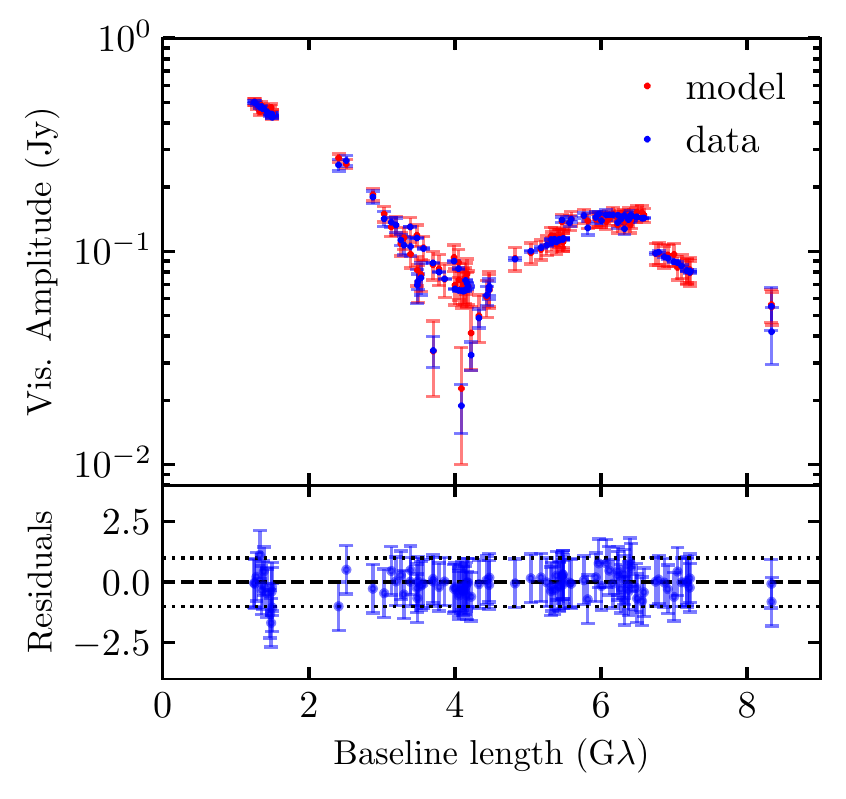}
\includegraphics[width=\columnwidth]{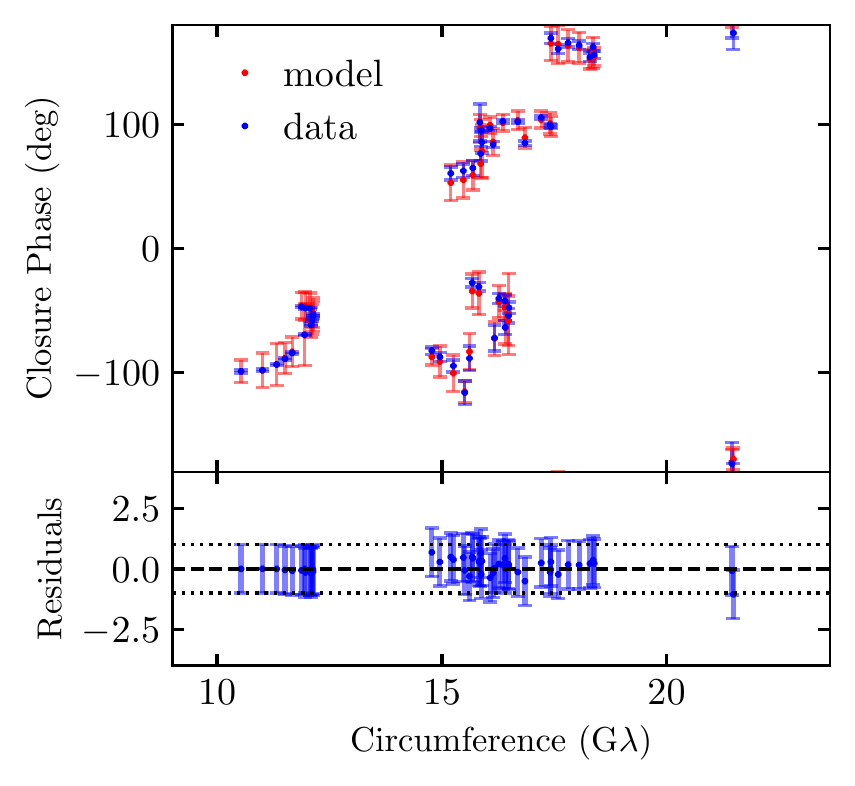}
\caption{{\it (Left)} Synthetic visibility amplitude data and the amplitudes that correspond to the most likely \texttt{PRIMO} reconstruction. The theoretical uncertainty (see section~\ref{sec:theory_err}) is shown as error bars on the model. The residuals of the fit are shown below, with the theoretical and observational uncertainties having been added in quadrature. {\it (Right)} The same but for the synthetic closure phase data.}
\label{fig:M47amp_clos}
\end{figure*}

Since the reconstructions with only 20 PCA components achieve errors which are comparable to the errors in the EHT 2017 data for M87, in this work we settle on fitting 20 PCA components to synthetic data as a proof of concept. However, a slightly higher or lower number of components may achieve comparable, or even better results. We use the results presented in Figure~\ref{fig:theory_err} to add a ``theoretical error'' to our model, which is implemented as an additional uncertainty, as a function of baseline length. In order to account for the fact that the peaks in the theoretical uncertainties shown in Figure~\ref{fig:theory_err} correspond to the locations of the visibility minima, which themselves scale inversely with $\theta_{\rm g}$, we scale the baseline lengths of the theoretical error curves in a similar way. Moreover, because the errors shown in this figure are fractional, we multiply them by the total flux $F$ in the image.

\subsection{Preparing Simulated Data}\label{sec:sim_data}

The EHT observations are simulated as follows. For each data point in the M87 EHT data, we use sinc interpolation to interpolate between pixels in $u-v$ space and approximate the visibility at that $u-v$ location. In order to mimic thermal noise, we dither each data point with errors derived from a Gaussian distribution with a standard deviation set by the error in the EHT data at each $u-v$ location for the 2017 EHT observations of M87. We do not include gain errors in our synthetic data at this time, nor do we include gains as free parameters in our model. 

\begin{figure}[t!]
\centering
\includegraphics[width=\columnwidth]{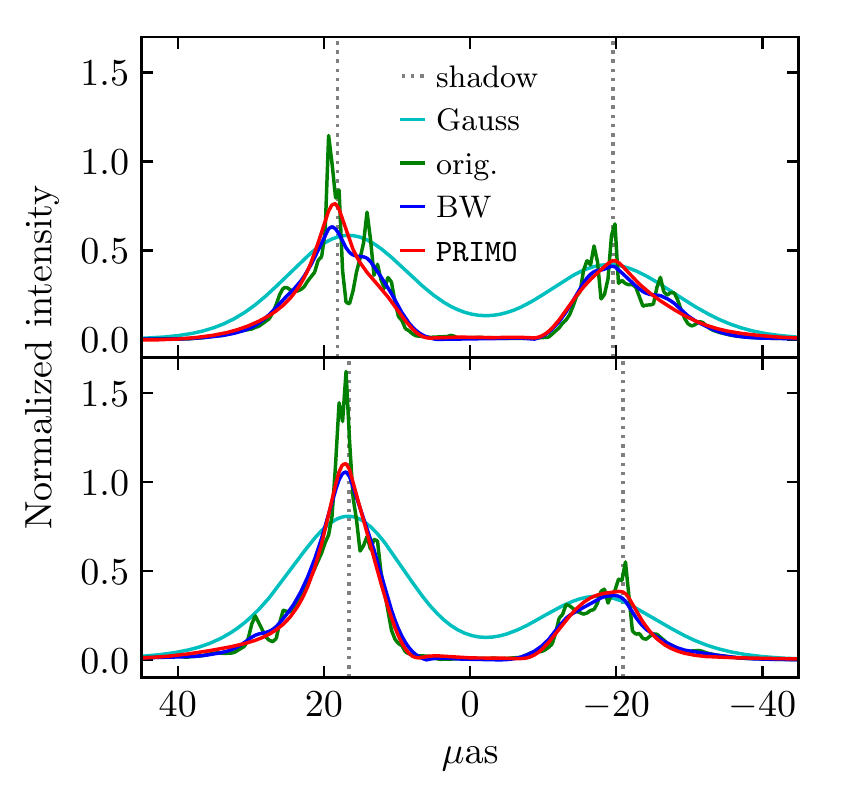}
\caption{Comparison of the horizontal ({\it top}) and vertical ({\it bottom}) cross sections of the images shown in Figure~\ref{fig:M47}. The curves show  the original snapshot (green), the snapshot filtered with a Butterworth filter (blue), the most-likely \texttt{PRIMO} reconstruction (red), the snapshot filtered with a 15 $\mu$as Gaussian (cyan), and the analytically calculated edges of the black hole shadow (gray dotted vertical lines). The cross sections are normalized such that all images have the same total flux, except for the Gaussian broadened image, which has 1.5 times the flux of the other images. The $y-$axis is in arbitrary units. \texttt{PRIMO} can accurately reproduce the main features of the image and does not introduce a significant bias in the ring size.}
\label{fig:M47cross}
\end{figure}

\section{Results from synthetic data}\label{sec:results}
\begin{figure*}[p]
\centering
\includegraphics[width=\textwidth]{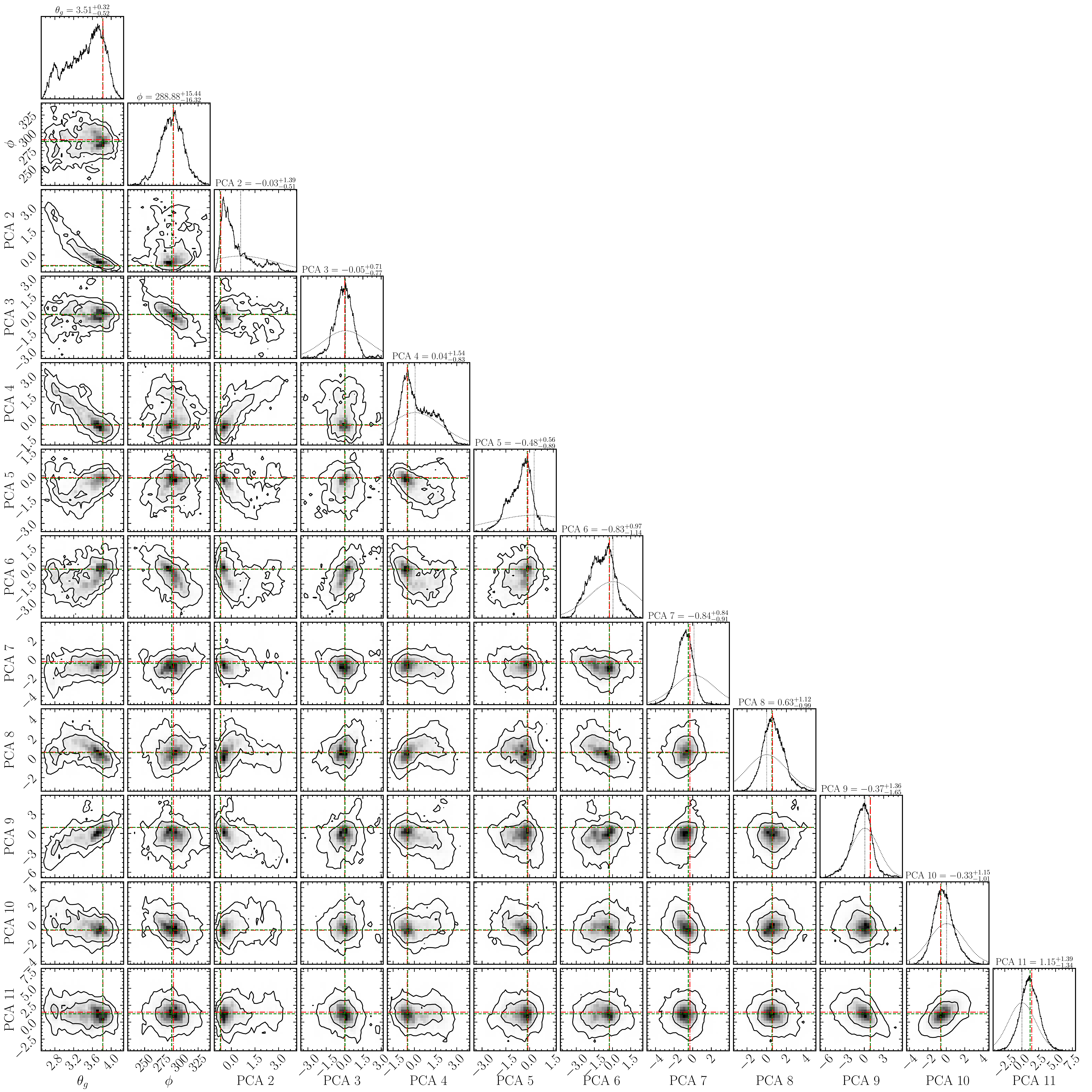}
\caption{Corner plot for the MCMC run that corresponds to  Figures~\ref{fig:M47}, \ref{fig:M47amp_clos}, and \ref{fig:M47cross}. Here, for brevity, we only include the flux parameter, the scaling parameter ($\theta_g$), the orientation parameter ($\phi$), and the amplitudes of the first 10 PCA components. The red vertical and horizontal dashed lines denote the highest-likelihood values while the vertical and horizontal green dashed lines denote the ground-truth values of these parameters. The dotted Gaussian curves and dotted vertical lines in the diagonal plots for the PCA amplitudes indicate the Gaussian prior for each amplitude used in the MCMC chain.}
\label{fig:M47corner}
\end{figure*}

In order to demonstrate the performance of \texttt{PRIMO} with EHT data, we apply it to a number of synthetic data sets created from simulated snapshots. We start with two snapshots from a single GRMHD+radiative transfer MAD simulation with electron number density scale $n_e = 10^5\,\mathrm{cm}^{-3}$, electron temperature parameter $R_{\mathrm{high}}=20$, black hole spin $a_{\mathrm{BH}}=0.9$ and  mass $M=6.5\times 10^9\,\mathrm{M}_{\odot}$. This set of parameters is relevant to the black hole in M87 and is consistent with the recent EHT results that showed that the polarization structure of M87 shows preference to MAD models over SANE models \citep{2021ApJ...910L..12E,2021ApJ...910L..13E}. These two snapshots were also considered in \citet{2022ApJ...928...55P} but for different values of the $R_{\rm high}$ parameter.

We begin by applying our algorithm to a simulated image snapshot that resembles a crescent shape but has some extended structure. This snapshot was not easily fit by a simple geometric crescent model \citep[][see Figures~16 and 17]{2022ApJ...928...55P}.  The top row of Figure~\ref{fig:M47} shows the simulated image and the highest likelihood reconstruction from \texttt{PRIMO} after 10,000,000 MCMC steps.
Unlike the geometric crescent model, \texttt{PRIMO} can easily reproduce the morphology of this image, arriving at the correct ring size and width, and the correct position angle for the peak of emission along the ring. 

 The bottom row of the Figure shows the original image blurred by a Gaussian filter with a width of $15\,\mu\mathrm{as}$ and the original image after it was filtered with an $n=2$, $r=15\,\mathrm{G}\lambda$ Butterworth filter.  The Gaussian broadened image approximates previously published EHT images, since most of the EHT reconstructed images published to date have been broadened by Gaussians. The width of the Gaussian kernel was chosen such that the median FWHM of the image, along 128 equispaced radial cross sections emanating from the center of the black-hole shadow, is equal to 20$\,\mu$as, i.e., similar to the M87 images reconstructed with other algorithms. (We note that we simply broadened the original simulated image and did not simulate CLEAN or RML imaging of it; still, the Gaussian broadened GRMHD image provides a simplified comparison to the resolution of previously published EHT images.) \texttt{PRIMO} achieves much higher image fidelity than the Gaussian blurred image and approaches the fidelity of the GRMHD input image simply blurred by the Butterworth filter.

Figure~\ref{fig:M47amp_clos} compares the visibility amplitudes and the closure phases of the synthetic data created from the simulated image as described in Section~\ref{sec:sim_data} to those of the reconstructed image with the highest likelihood. The model shows very good agreement with the synthetic data and no structure is present in the residuals. As expected, because of the very large signal-to-noise ratio of most of the EHT measurements, the residuals are dominated by the theoretical errors introduced by the truncation in the number of PCA components used. Nevertheless, this truncation does not introduce any substantial biases in the image structure or its properties.

Figure~\ref{fig:M47cross} compares the vertical (N-S) and horizontal (E-W) cross sections of the original image, the Butterworth filtered snapshot, the Gaussian filtered snapshot, and the most likely reconstruction with \texttt{PRIMO}. There is remarkable agreement between the properties of the reconstrcted image and those of the original one. In particular, the \texttt{PRIMO} fit is a much more accurate representation of the original snapshot than the snapshot convolved with a 20$\,\mu$as beam. The main features of the cross sections, i.e., the location and amplitude of the peaks, the width of the peaks, the size and depth of the central flux depression, and the relative amplitude difference between the two peaks is well approximated by the reconstruction.

\begin{figure*}[t!]
\centering
\includegraphics[width=\textwidth]{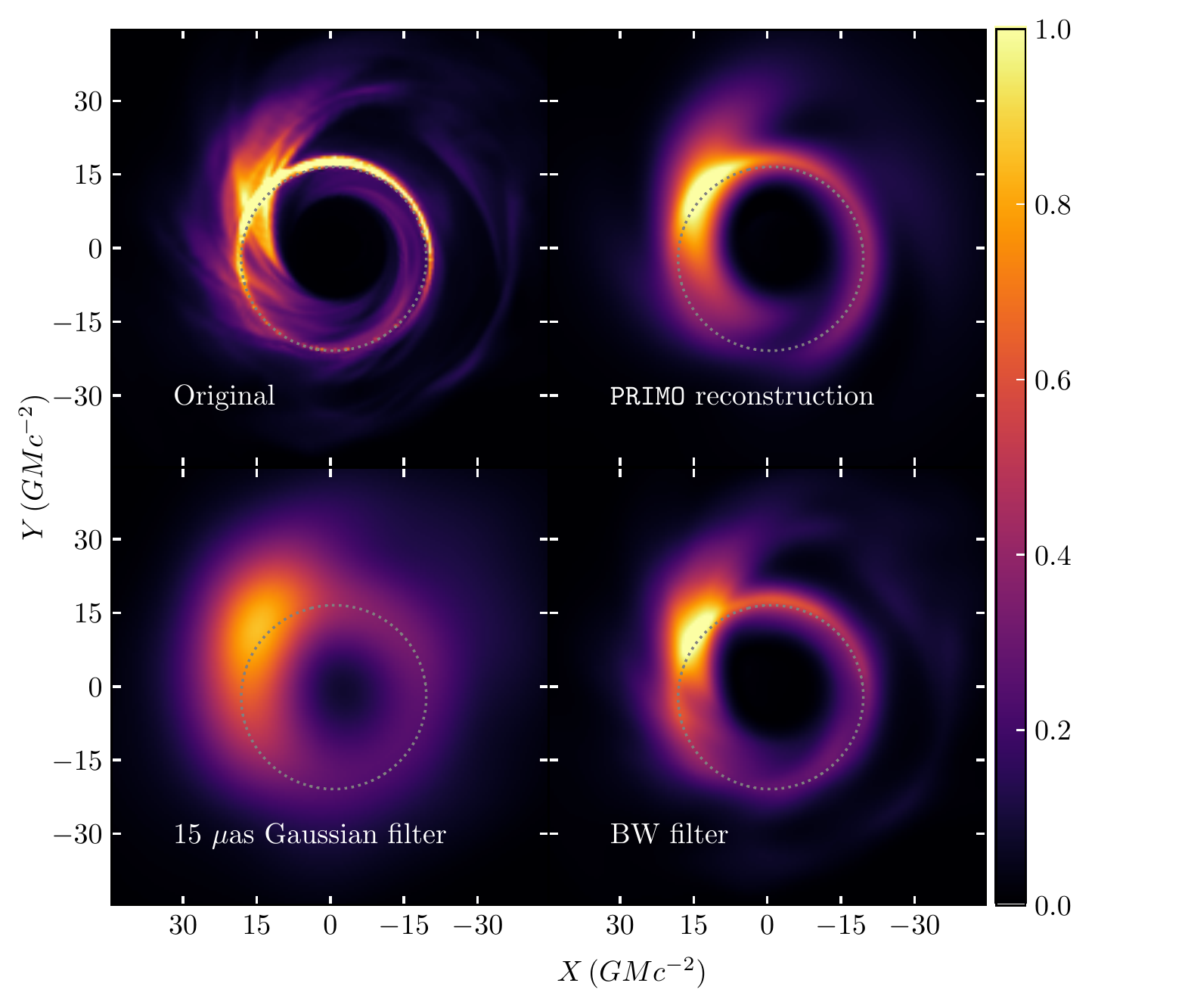}
\caption{Same as Figure~\ref{fig:M47} but for the second synthetic data set we consider. }
\label{fig:M191}
\end{figure*}

\begin{figure*}[t!]
\centering
\includegraphics[width=\columnwidth]{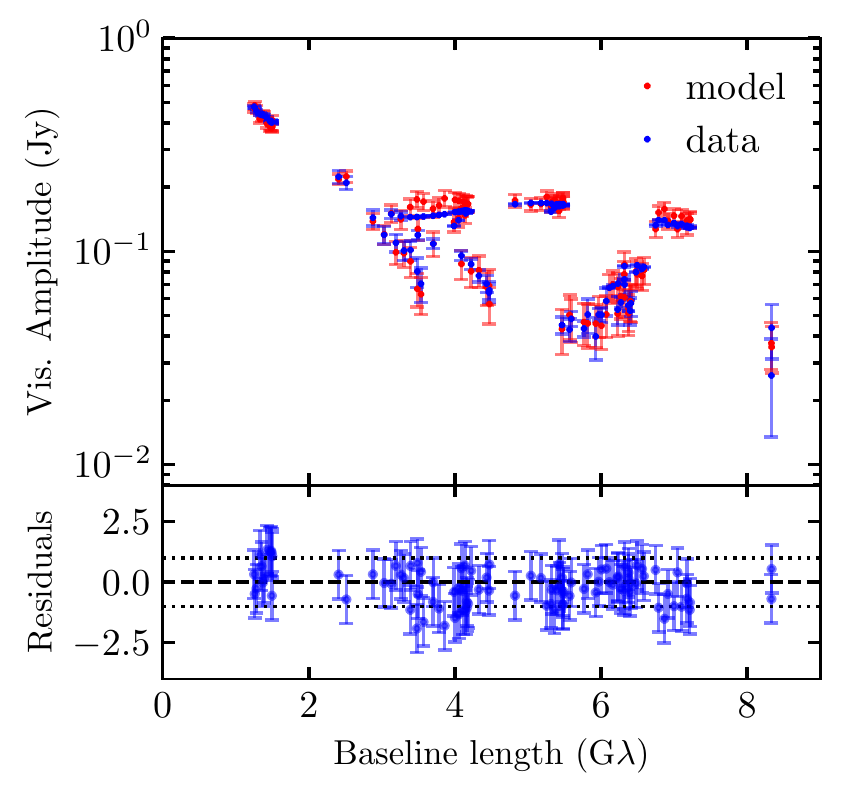}
\includegraphics[width=\columnwidth]{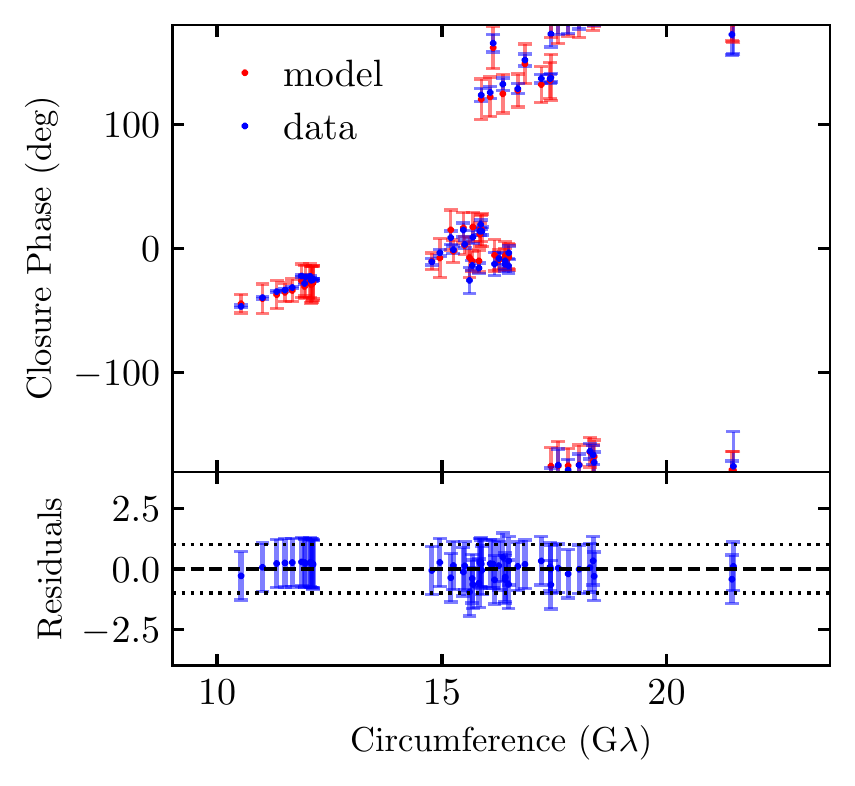}
\caption{Same as Figure~\ref{fig:M47amp_clos} but for the same synthetic data set as shown in  Figure~\ref{fig:M191}.}
\label{fig:M191amp_clos}
\end{figure*}

Figure~\ref{fig:M47corner} shows a corner plot for numerous key parameters for the MCMC run discussed above. The corner plot shows a few correlations between parameters, such as between the scaling parameter $\theta_g$ and the amplitude of the second PCA component, as well as with several other components but to a lesser extent. Although the PCA components are orthogonal when considered across the entire image (or $u-v$ space), they are no longer orthogonal when we consider only the discrete locations of the EHT baselines. Because of this, some correlations between different PCA components are also visible, such as between the second and fourth components. The correlation between the overall scale ($\theta_g$) and the second component is not surprising; the second component affects the width of the ring, which is highly correlated with the diameter of the ring. 

The widths of the posteriors of most of the low-order PCA components are significantly smaller than those of the priors, demonstrating that the broad-brush properties of the reconstructed image are driven by the data and not by the priors. This is increasingly less the case for the higher-order PCA components, justifying the level at which we truncated the series of components. The Figure also compares the ground-truth values (shown in green) to the highest likelihood values from the reconstruction (shown in red)\footnote{For the amplitudes of the PCA components, we treat the amplitudes derived by projecting the original image onto the first $N$ PCA components as the ground-truth values.}. In all cases, there is a remarkable agreement between the two.

\begin{figure}[t!]
\centering
\includegraphics[width=\columnwidth]{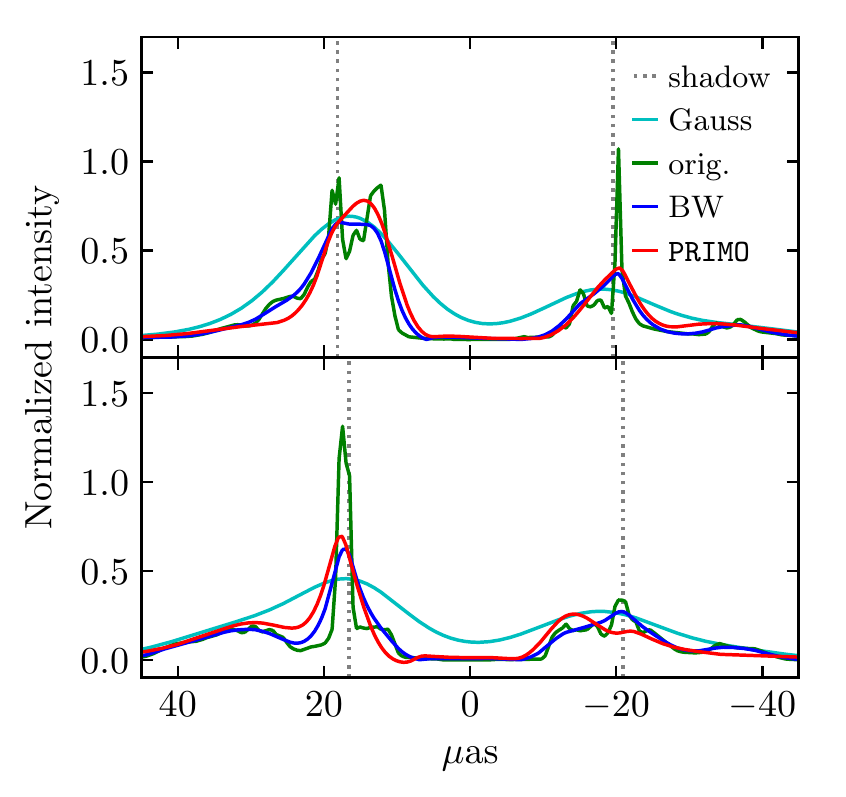}
\caption{Same as Figure~\ref{fig:M47cross} but for the second synthetic data set.}
\label{fig:M191cross}
\end{figure}

\begin{figure*}[p]
\centering
\includegraphics[width=\textwidth]{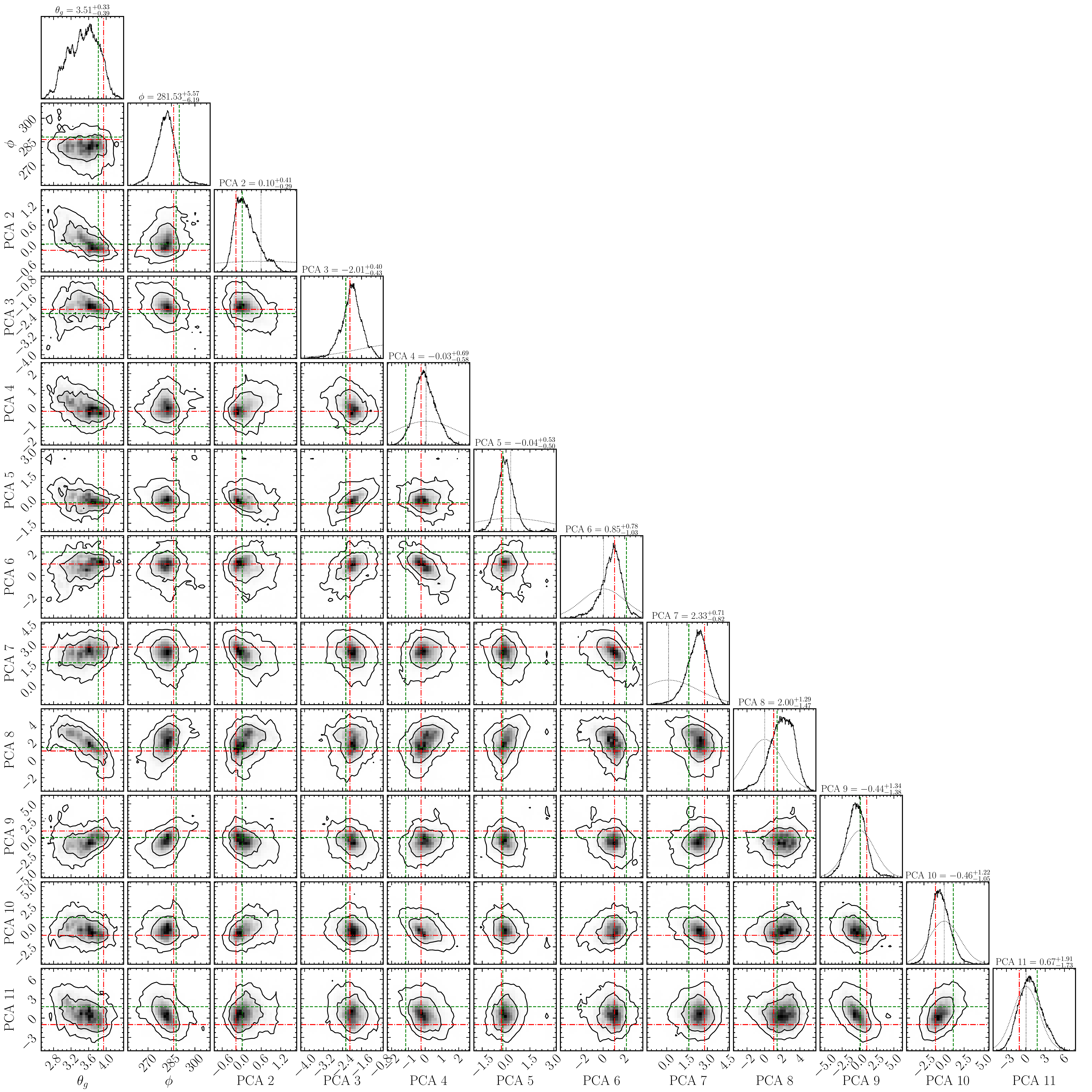}
\caption{Same as Figure~\ref{fig:M47corner} but for the MCMC run shown in Figures~\ref{fig:M191}, \ref{fig:M191amp_clos} and \ref{fig:M191cross}.}
\label{fig:M191corner}
\end{figure*}

As a second example, we apply \texttt{PRIMO} to synthetic data generated from a second snapshot that is dominated by an extended flux tube. The geometric crescent fit to this image failed to reconstruct a reasonable ring size even when a Gaussian component was added to the model (see Figures~18 and 19 in \citealt{2022ApJ...928...55P}). However, as can be seen in Figures~\ref{fig:M191}-\ref{fig:M191corner}, \texttt{PRIMO} can accurately reconstruct the location of the peak of emission along the ring, the width of the peak, the shape and depth of the central flux depression, and the extended flux tube towards the top left of the image. The visibility amplitudes and closure phases from the reconstructed image show good agreement with the synthetic data and very little structure is visible in the residuals. 

\begin{figure*}[t!]
\centering
\includegraphics[width=\textwidth]{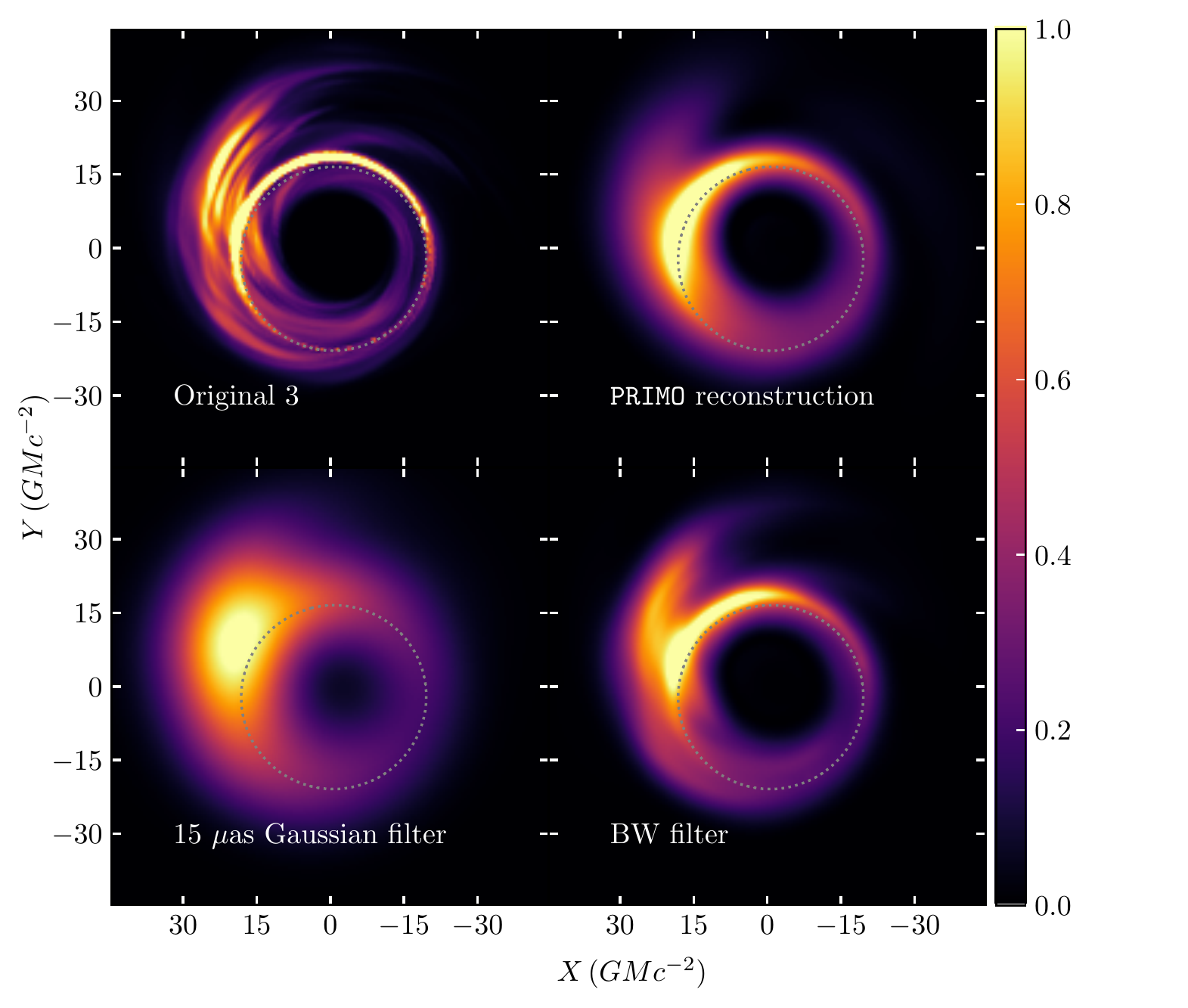}
\caption{Same as Figure~\ref{fig:M47} but for the third synthetic data set, which was not contained in the training set but was generated from a GRMHD simulation with a SANE magnetic field geometry and a black-hole spin of $a_{\mathrm{BH}}=0.7$. Despite not being part of the training set, salient features of the images are accurately reconstructed by \texttt{PRIMO}.}
\label{fig:S7314}
\end{figure*}

\begin{figure*}[t!]
\centering
\includegraphics[width=\columnwidth]{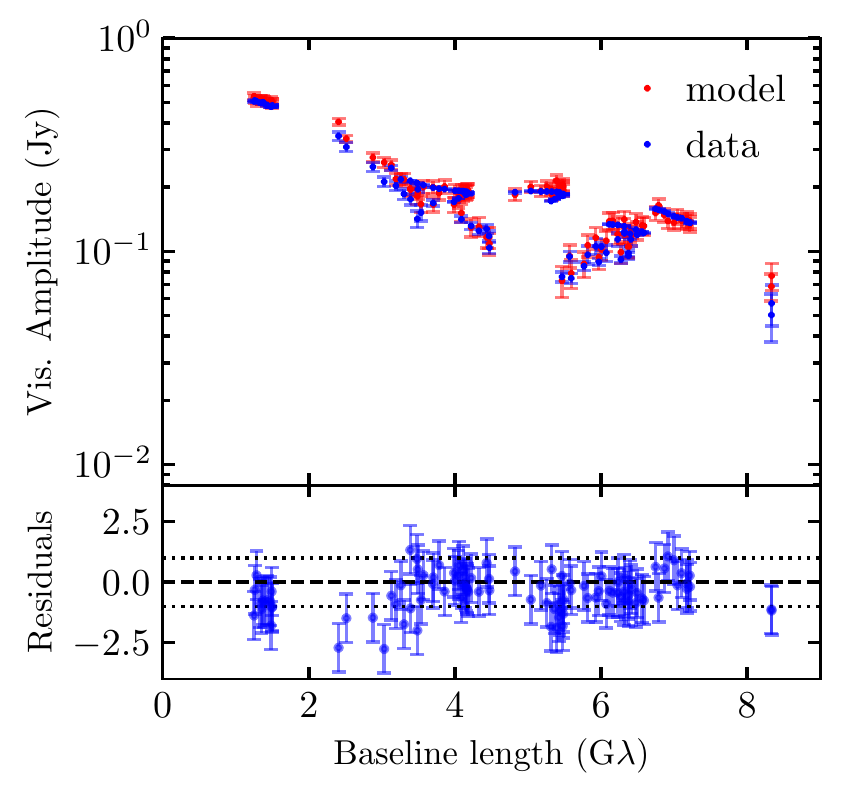}
\includegraphics[width=\columnwidth]{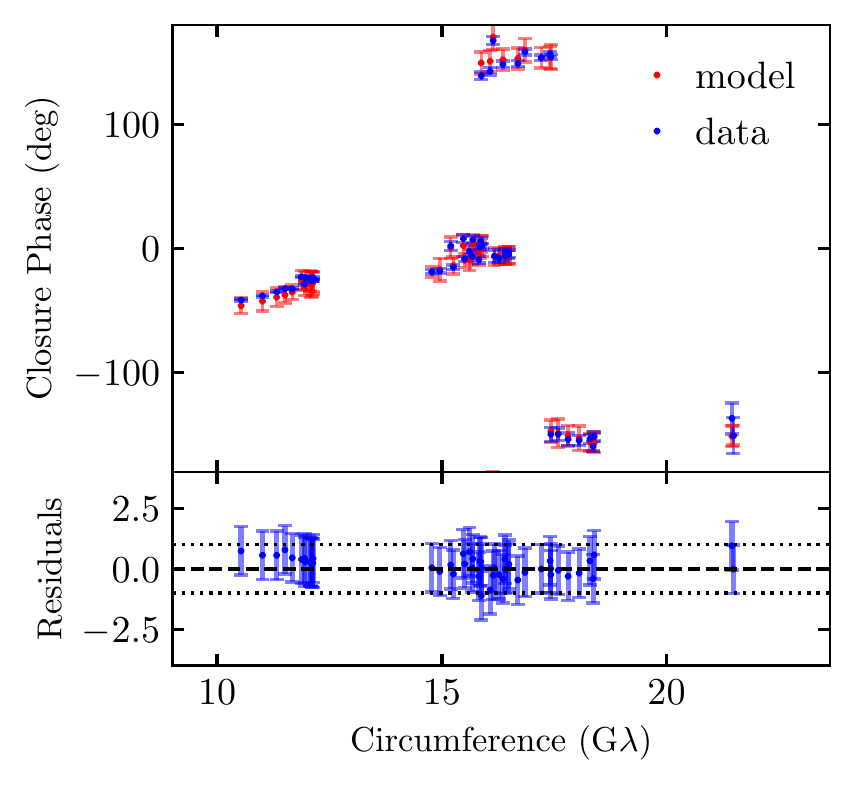}
\caption{Same as Figure~\ref{fig:M47} but for the third synthetic data set we consider.}
\label{fig:S7314amp_clos}
\end{figure*}

\begin{figure}[t!]
\centering
\includegraphics[width=\columnwidth]{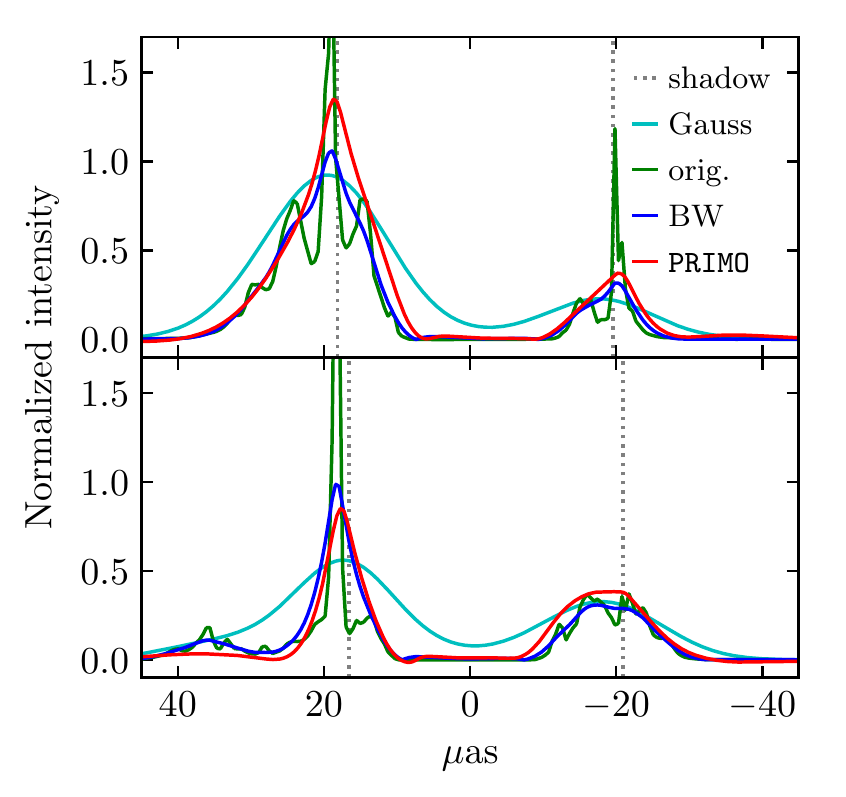}
\caption{Same as Figure~\ref{fig:M47cross} but for the third synthetic data set we consider. }
\label{fig:S7314cross}
\end{figure}

Finally, we consider an image that is not included in the training ensemble used to generate the PCA components. While all images in the training set have a black hole spin of $a_{\mathrm{BH}}=0.9$, for the final synthetic data set we use an image from a simulation with a black hole spin of $a_{\mathrm{BH}}=0.7$. This image has a SANE magnetic field geometry and a plasma parameter of $R_{\mathrm{high}}=20$. Figures~\ref{fig:S7314}, \ref{fig:S7314amp_clos}, and \ref{fig:S7314cross} show the results of reconstructing this image with \texttt{PRIMO}. Even though this image was not included in the ensemble used to generate the PCA components, the algorithm was still able to accurately reconstruct the salient image features, such as the depth and shape of the brightness depression, the size and width of the peak, the orientation of the peak brightness asymmetry in the ring feature, and the extended structure towards the top left of the image. 

\section{Summary}\label{sec:discussion}

We have presented a novel PCA-based image reconstruction algorithm, \texttt{PRIMO}, for reconstruction of black hole images from EHT data. Our algorithm is unique in that it combines prior information from physically motivated simulations to reconstruct images that lie in the same general space of images spanned by the simulations. Each simulation can create countless images with different morphologies due to the turbulent nature of the accretion flow, making it unlikely that the particular realization of the turbulent flow of the source that the EHT observes would be well fit by any one of the thousands of simulation images included in our library. However, the PCA-based algorithm allows us to reconstruct images regardless of whether or not they are contained within the library of images from which the PCA basis was created. Compared to the results of previous work, \texttt{PRIMO} is not severely affected by the biases identified in \citet{2022ApJ...928...55P}, where simulated images were fit with analytic crescent models.

Throughout this work we have used the EHT baseline coverage from the 2017 observations. Since then, the EHT has observed several more times with additional telescopes. We expect that, with additional baselines, we will be able to incorporate a higher number of PCA components to generate images from the data and achieve even better angular resolution. The EHT is also planning to observe at 345\,GHz in the coming years, which will allow us to probe even higher spatial frequencies. \texttt{PRIMO} can easily be adapted to exploit these new observations. 

\begin{acknowledgements}
We thank C. K. Chan, P. Hallur, and B. Zackay for useful discussions. L.\;M.\ gratefully acknowledges support from an NSF Astronomy and Astrophysics Postdoctoral Fellowship under award no. AST-1903847. D.\;P.\, and F.\;O.\, gratefully acknowledge support from NSF PIRE grant 1743747 for this work. All ray tracing calculations were performed with the \texttt{El~Gato} GPU cluster at the University of Arizona that is funded by NSF award 1228509. 
\end{acknowledgements}


\bibliography{main}
\end{document}